\documentclass[showpacs,preprintnumbers,amsmath,amssymb,floatfix]{revtex4}

\usepackage{graphicx}
\usepackage{epsfig}
\usepackage{amsfonts}
\usepackage{amssymb}
\usepackage{epsf}
\newcommand{\insertplot}[5]{\begin{figure}
 \hfill\hbox to 0.05in{\vbox to #5in{\vfill
 \inputplot{#1}{#4}{#5}}\hfill}
 \hfill\vspace{-.1in}
 \caption{#2}\label{#3}
 \end{figure}}
\newcommand{\inputplot}[3]{
 \special{ps: plotfile #1}
}

\newcounter{fig}   

\newcommand{\vphi}{\varphi}

\begin{document}

\title{
Rapidly Rotating Neutron Stars in Dilatonic Einstein-Gauss-Bonnet Theory}

\author{
{\bf Burkhard Kleihaus$^1$, Jutta Kunz$^1$, Sindy Mojica$^1$,
and Marco Zagermann$^2$}
}
\affiliation{
$^1${Institut f\"ur Physik, Universit\"at Oldenburg,
D-26111 Oldenburg, Germany}\\
$^2${Institut f\"ur Theoretische Physik, Leibniz Universit\"at Hannover,
D-30176 Hannover, Germany}
}
\date{\today}
\begin{abstract}
We construct sequences of rapidly rotating neutron stars
in dilatonic Einstein-Gauss-Bonnet theory,
employing two equations of state for the nuclear matter.
We analyze the dependence of the physical properties of these
neutron stars on the Gauss-Bonnet coupling strength.
For a given equation of state we determine the physically relevant domain
of rapidly rotating neutron stars, which is delimited by
the set of neutron stars rotating at the Kepler limit,
the set of neutron stars along the secular instability line,
and the set of static neutron stars.
As compared to Einstein gravity, the presence of
the Gauss-Bonnet term decreases this domain,
leading to lower values for the maximum mass 
as well as to smaller central densities.
The quadrupole moment is decreased by the Gauss-Bonnet term 
for rapidly rotating neutron stars, 
while it is increased for slowly rotating neutron stars.
The universal relation between the quadrupole moment 
and the moment of inertia found in General Relativity
appears to extend to dilatonic Einstein-Gauss-Bonnet theory
with very little dependence on the coupling strength 
of the Gauss-Bonnet term.
The neutron stars carry a small dilaton charge.
\end{abstract}
\maketitle

\section{Introduction}

Neutron stars 
represent the compact remains of massive stars
after their supernova explosion and collapse.
%
Consisting largely of neutron matter, neutron stars
form highly compact astrophysical objects. 
Therefore it is essential 
to take into account the curvature of space-time
induced by the large concentration of mass.
Coupling the nuclear matter to gravity
as described by General Relativity (GR)
or generalized theories of gravity
then leads to a consistent theoretical approach
to study neutron stars and their properties.

A current major unknown in such studies is 
the proper treatment of nuclear matter under extreme
conditions as encountered inside a neutron star.
Here numerous equations of state (EOSs)
for the nuclear matter have been proposed and employed
(see e.g.~\cite{Lattimer:2012nd}).
The choice of the EOS
determines the size of the neutrons stars
and the maximal value of the mass.
The observations of neutron stars with masses of $M \approx 2 M_\odot$
\cite{Demorest:2010bx,Antoniadis:2013pzd}
therefore provide a strong constraint for the physically viable EOSs.

While the physical properties of neutron stars 
typically depend strongly on the chosen EOS,
in recent years
the study of EOS independent -- or almost independent --
neutron star characteristics came into the focus. 
Notably, the $I$-Love $Q$ relations for neutron stars
represent universal relations that hold between the scaled moment
of inertia, the Love number, and the scaled quadrupole moment
in Einstein gravity \cite{Yagi:2013bca,Yagi:2014bxa,Yagi:2014qua}.
Also for quasinormal modes (QNMs) of neutron stars universal relations
concerning frequency and damping time have been found 
\cite{Andersson:1997rn,Benhar:1998au,BlazquezSalcedo:2012pd,Blazquez-Salcedo:2013jka}.

Whereas most studies of neutron star properties have been performed within GR,
generalized models of gravity have also been considered
(for a recent review see \cite{Berti:2015itd}).
In fact, neutron stars represent an excellent testing ground for such
generalized models of gravity.
While scalar-tensor theories (STTs), for instance, can lead to
results very close to those of GR, certain STTs also
allow for the phenomenon of spontaneous scalarization,
yielding neutron stars with considerably larger masses
in the presence of a non-trivial scalar field
\cite{Damour:1993hw,Doneva:2013qva}.

Here we consider neutron stars in dilatonic Einstein-Gauss-Bonnet (dEGB) theory.
This theory is motivated from string theory, a leading candidate
for a quantum theory of gravity and a unified 
description of the fundamental interactions of Nature.
String theory predicts the presence of higher curvature terms in the action
as well as further fields.
In particular, 
the low energy effective action obtained from heterotic
string theory contains as basic ingredients a Gauss-Bonnet (GB)
term and a dilaton field \cite{Gross:1986mw,Metsaev:1987zx}.
As an attractive feature of dEGB theory 
the quadratic curvature terms in the action 
still lead to only second order equations of motion.

Properties of static neutron stars in dEGB theory were first
considered in \cite{Pani:2011xm}.
Studying neutron stars for three EOSs it was shown, 
that the maximum mass of the
neutron stars decreases as the GB coupling constant is increased.
Interestingly, the sequences of static neutron stars cannot 
be extended beyond a critical
value of the central density, which depends on the GB coupling strength
and on the EOS.
Here a vanishing radicant is encountered in the expansion of the dilaton field
at the origin \cite{footnote1}.
This behavior therefore leads to EOS dependent constraints
on the GB coupling \cite{Pani:2011xm}.

In the case of slow rotation a perturbative study allows 
for the extraction of the moment of inertia of neutron stars.
Generalizing the corresponding GR derivation \cite{Hartle:1967he},
the moment of inertia was calculated in dEGB theory in
\cite{Pani:2011xm}. There it was shown that the moment of inertia 
decreases with increasing GB coupling.
Recently, also QNMs of neutron stars were studied
in dEGB theory \cite{Blazquez-Salcedo:2015ets},
showing that the frequency of the modes is increased by 
the presence of the GB term.

Rapidly rotating neutron stars have been studied extensively in GR 
(see e.g.~\cite{Stergioulas:2003yp,Gourgoulhon:2010ju,Friedman:2013xza}).
However, in generalized theories of gravity the investigation of 
rapidly rotating neutron stars has only begun recently
(see e.g.~\cite{Berti:2015itd}),
where much progress was achieved in STTs \cite{Doneva:2013qva}
and first results were obtained in dEGB theory \cite{Kleihaus:2014lba}.
The physically relevant domain of neutron stars is delimited by the set of static neutron stars,
by the set of neutron stars along the secular instability line,
which possess maximum mass for fixed angular momentum,
and by the set of neutron stars rotating at the Kepler limit.

Here we construct for the first time the full physically relevant
domain for rapidly rotating neutron stars in dEGB theory, 
employing two well-known EOSs 
\cite{Diaz-Alonso:1985,Lorenz:1992zz,Haensel:2004nu}.
We discuss the dependence of the mass on the radius and on the central density.
We consider the compactness, the angular momentum, the rotation
period and the dilaton charge. We extract the moment of inertia
and the quadrupole moment of these neutron stars,
and analyze the corresponding universal relation.
Finally we discuss the deformation of the rapidly rotation neutron stars.
For all these physical properties we analyze the dependence 
on the GB coupling strength.

The paper is organized as follows:
In section II we exhibit the action, the field equations,
the Ans\"atze, the boundary conditions and the 
definitions of the physical quantities, 
while we present our main results on 
rapidly rotating neutron stars in dEGB theory in section III.
Section IV gives our conclusions and outlook.
The Appendix sketches our derivation of the quadrupole moment for neutron stars.

\section{Einstein-Gauss-Bonnet-Dilaton Theory}

Here we first motivate and recall the action of dEGB theory.
We exhibit the equations of motion for the metric and the 
dilaton field as well as the constraint equations for
the stress energy tensor. Subsequently, we present the
stationary axially symmetric Ansatz for the metric. 
For the neutron star matter we assume a perfect fluid 
in uniform rotation, described by a polytropic EOS. 
Since we work with dimensionless coordinates
we discuss our choice of the dimensionful scales
to obtain the corresponding physical values of the
observables. We recall how to extract the global charges and the quadrupole moment
 from the asymptotic expansions of the 
functions, and we describe how we analyze the size and the shape of the
neutron stars.

\subsection{Action and Field Equations}

Today string theory represents a promising approach
towards quantum gravity and a
unified description of the fundamental interactions.
In string theory modifications of GR arise, which 
can be incorporated into an effective low energy action.
In particular, in heterotic string theory
a Gauss-Bonnet term arises, which is coupled to
a modulus field, the dilaton
\cite{Gross:1986mw,Metsaev:1987zx}.
Moreover, Lorentz-Chern-Simons terms and Kalb-Ramond axions
are present.
(For more recent discussions on the low energy effective action and its maximally symmetric solutions
see e.g.~\cite{Chemissany:2007he,Lechtenfeld:2010dr,Green:2011cn,Gautason:2012tb,Quigley:2015jia, Kutasov:2015eba}.)

The low energy effective action has received much attention in 
connection with black holes. 
Based on the inclusion of various parts of the effective action,
numerous static and slowly rotating black hole solutions have been found
(see e.g.
\cite{Gibbons:1982ih,Gibbons:1987ps,Callan:1988hs,Campbell:1990ai,Garfinkle:1990qj,
Campbell:1991rz,Shapere:1991ta,Campbell:1991kz,Mignemi:1992nt,Mignemi:1992pm,
Horne:1992bi,Kanti:1995cp}),
as well as rapidly rotating ones
(see e.g.~\cite{Frolov:1987rj,Sen:1992ua,Garcia:1995qz,Rasheed:1995zv,Youm:1997hw}).
Also string theory corrections of further
compact objects have been considered
(see e.g.~\cite{Berti:2015itd}).

Motivated by the low-energy heterotic string theory action
\cite{Gross:1986mw,Metsaev:1987zx},
we here employ a certain simplified action, which has been 
considered previously for black holes
\cite{Mignemi:1992nt,Kanti:1995vq,Torii:1996yi,Alexeev:1996vs,Moura:2006pz,
Guo:2008hf,Chen:2009rv,Pani:2009wy,Pani:2011gy,
Kleihaus:2011tg,Ayzenberg:2014aka,Maselli:2015tta,Kleihaus:2015aje}
and wormholes \cite{Kanti:2011jz}.
In particular, this action retains only the dilaton and the GB term
in addition to the Einstein-Hilbert action, while it neglects
the Lorentz-Chern-Simons and axion terms (as well as gauge fields and possible matter-dilaton couplings)
and treats spacetime as effectively four-dimensional, assuming no light compactification moduli. 
Thus the action reads
\begin{eqnarray}  
S=\frac{c^4}{16\pi G}\int d^4x \sqrt{-g} \left[R - \frac{1}{2}
 \partial_\mu \phi \,\partial^\mu \phi
 + \alpha  e^{-\gamma \phi} R^2_{\rm GB}   \right] + S_{\rm matter} ,
\label{act}
\end{eqnarray} 
where 
$\phi$ denotes the dilaton field
with coupling constant $\gamma$, 
$\alpha$ is a positive  
coefficient given in terms of the Regge slope parameter, $\alpha'$, as $\alpha=\alpha'/8$ 
and
$R^2_{\rm GB} = R_{\mu\nu\rho\sigma} R^{\mu\nu\rho\sigma}
- 4 R_{\mu\nu} R^{\mu\nu} + R^2$ 
represents the GB term. 
With $S_{\rm matter}$ we indicate the action of the nuclear matter,
although we here construct the neutron stars by assuming a perfect fluid
with a given EOS. We should also note that even though our approach is string inspired, we would like to be as general as possible and do
not assume any a priori restriction on the parameter $\alpha$ (e.g. from connecting string theory with elementary particle physics data) other than direct empirical constraints from 
astronomical observations that we will recall at the beginning of section III. 

Variation of the action then leads to a coupled set of equations,
to be solved subject to certain boundary conditions and constraints.
The dilaton and the generalized Einstein equations are given by
\begin{eqnarray}
\nabla^2 \phi & = & \alpha \gamma  e^{-\gamma \phi}R^2_{\rm GB}
\label{dileq}\\
G_{\mu\nu} & = &
\frac{1}{2}\left[\nabla_\mu \phi \nabla_\nu \phi 
                 -\frac{1}{2}g_{\mu\nu}\nabla_\lambda \phi \nabla^\lambda\phi 
		 \right]
\nonumber\\
& &
-\alpha e^{-\gamma \phi} 
\left[	H_{\mu\nu}
  +4\left(\gamma^2\nabla^\rho \phi \nabla^\sigma \phi
           -\gamma \nabla^\rho\nabla^\sigma \phi\right)	P_{\mu\rho\nu\sigma}
		 \right]
\nonumber\\
& &
		 +8\pi \beta T_{\mu\nu}
\label{Einsteq}
\end{eqnarray}
with
\begin{eqnarray}
H_{\mu\nu} & = & 2\left[R R_{\mu\nu} -2 R_{\mu\rho}R^\rho_\nu
                        -2 R_{\mu\rho\nu\sigma}R^{\rho\sigma}
			+R_{\mu\rho\sigma\lambda}R_\nu^{\ \rho\sigma\lambda}
		   \right]
		   -\frac{1}{2}g_{\mu\nu}R^2_{\rm GB}	\ ,
\\
 P_{\mu\nu\rho\sigma} & = & 
R_{\mu\nu\rho\sigma}
+2 g_{\mu [ \sigma} R_{\rho ]\nu}
+2 g_{\nu [ \rho} R_{\sigma ]\mu}
+R g_{\mu [ \rho} g_{\sigma ]\nu} \ ,
\label{HP}
\end{eqnarray}
and coupling constant $\beta= G/c^4$.

Here we have introduced on the r.h.s.~of the generalized Einstein
equations the stress energy tensor of the neutron star matter
in the form of a perfect fluid
\begin{equation}
T_{\mu\nu} = \frac{1}{c^2}\left(\epsilon + P\right) U_\mu U_\nu + P g_{\mu\nu} \ ,
\label{tmunu}
\end{equation}
where $\epsilon$ and $P$ denote the energy density and the pressure of
the neutron star fluid, respectively,
and $U_\mu$ represents the four velocity of the fluid.
In order to close the system of PDEs we impose the condition
that the stress energy tensor is covariantly conserved,
\begin{equation}
\nabla_\mu T^{\mu\nu} = 0 \ .
\label{Tpde}
\end{equation}

\subsection{Ans\"atze for the Metric and the Fluid}

To obtain rotating neutron stars
we employ the Lewis-Papapetrou line element \cite{Wald:1984rg} 
for a stationary, axially symmetric spacetime with
two Killing vector fields $\xi=\partial_t$,
$\eta=\partial_\varphi$. 
In terms of the spherical coordinates $r$ and $\theta$, 
the quasi-isotropic metric then reads \cite{Kleihaus:2000kg}
\begin{equation}
ds^2 = g_{\mu\nu}dx^\mu dx^\nu
= -c^2 e^{2 \nu_0}dt^2
      +e^{2(\nu_1-\nu_0)}\left(e^{2 \nu_2}\left[dr^2+r^2 d\theta^2\right] 
       +r^2 \sin^2\theta
          \left(d\varphi-\omega  dt\right)^2\right) .
\label{metric} 
\end{equation}
The metric functions $\nu_0$, $\nu_1$, $\nu_2$ and $\omega$ 
as well as the dilaton function $\phi$ 
depend on the coordinates $r$ and $\theta$, only.

We here consider uniform rotation of the neutron star fluid,
an assumption well justified for most neutron stars 
\cite{Friedman:2013xza}.
In this case the four velocity has the form
\begin{equation}
U^\mu = \left( u, 0, 0, \Omega u\right) \ , 
\label{Umu}
\end{equation}
where $\Omega$ denotes the constant angular velocity of the star.

Before proceeding further, let us introduce the dimensionless quantities
\begin{equation}
\hat{r}=\frac{r}{r_0} \ , \ \ \
\hat{t}=\frac{t c}{r_0} \ , \ \ \
\hat{\omega}=\frac{\omega r_0}{c} \ , \ \ \
\hat{\Omega}=\frac{\Omega r_0}{c} \ , 
\end{equation}
and
\begin{equation}
\hat{\epsilon}=\frac{\epsilon}{\epsilon_0}\ , \ \ \
\hat{P}=\frac{P}{\epsilon_0}\ , \ \ \
\hat{T}_{\mu\nu}=\frac{T_{\mu\nu}}{\epsilon_0}\ ,
\end{equation}
with $r_0$ representing a length scale and $\epsilon_0$
an energy density.
Substitution of these expressions in Eqs.~(\ref{dileq}) and (\ref{Einsteq})
then suggests the introduction of the following 
dimensionless coupling constants
\begin{equation}
\hat{\alpha}=\frac{\alpha}{r_0^2} \ , \ \ \ 
\hat{\beta} =r_0^2 \epsilon_0 \beta =\frac{G r_0^2 \epsilon_0}{c^4} \ .
\label{ab}
\end{equation}

To fix the scales let us begin by considering the asymptotic behaviour 
of the function $\nu_0$
\begin{equation}
\nu_0 \approx \frac{GM}{c^2 r} = \frac{GM}{c^2 r_0 \hat{r}} 
= \frac{M}{M_0\hat{r}}
= \frac{\hat M}{\hat{r}} ,
\end{equation}
%
where $\hat M$ is the dimensionless mass.
Thus the mass scale $M_0$ is related to the length scale $r_0$
by $M_0 = r_0 c^2/G$. 
Subsequently we choose the dimensionless coupling constant $\hat{\beta}=1$. 
This then relates the energy density scale 
to the length scale by $\epsilon_0 =c^4/(G r_0^2)$.
With the choice $M_0=M_\odot$ we obtain
$r_0 = 1.476902$km, $\epsilon_0/c^2=617.394 \times 10^{15}{\rm g}/{\rm cm}^3$.
Finally, we rename the dimensionless quantities omitting the hat.

Let us now address the neutron star matter once more.
Employing the normalization condition for the four
velocity of the fluid $U^\mu U_\mu = -1$, the velocity function
$u$ can be expressed in terms of the metric functions 
$\nu_0$, $\nu_1$ and $\omega$ and the constant 
angular velocity $\Omega$,
\begin{equation}
u^2 = \frac{e^{-2\nu_0}}{1-(\Omega-\omega)^2 r^2\sin^2\theta e^{2\nu_1-4\nu_0}}  .
\label{u2}
\end{equation}

The constraints $\nabla_\mu T^{\mu\nu}= 0$
yield the differential equations for the pressure $P$ 
and the energy density $\epsilon$ 
\begin{equation}
\frac{\partial_r P}{\epsilon +P}  =  \frac{\partial_r u}{u} 
 , \ \ \
\frac{\partial_\theta P}{\epsilon +P}  =  \frac{\partial_\theta u}{u} .
\label{dpres}
\end{equation}
These equations have to be supplemented by an EOS, 
$\epsilon = \epsilon(P)$ (or $P = P(\epsilon)$).

A polytropic EOS relates the pressure $P$ to the  baryon mass density $\rho$
according to
\cite{Friedman:2013xza}
\begin{equation}
P=K \rho^{\Gamma} \ , \ \ \ \Gamma = 1 + \frac{1}{N}
\label{polEOSprho}
\end{equation}
with polytropic constant $K$, polytropic exponent $\Gamma$,
and polytropic index $N$,
while the energy density $\epsilon$ of a polytrope is given by
\begin{equation}
\epsilon = NP + \rho .
\end{equation}
It is common practice to parametrize the pressure and
the energy density by the function $\Theta$,
\begin{equation}
P = P_0 \Theta^{N+1}, \ \ \  \epsilon = \left(N P + \rho_0 \Theta^N \right)\ ,
\label{polEOS}
\end{equation}
where $P_0$ and $\rho_0$ are dimensionless constants.
Substitution of these expressions into Eq.~(\ref{dpres}) yields
\begin{equation}
\Theta = c_0 u - \frac{\rho_0}{P_0(N+1)} \ ,
\label{polThe}
\end{equation}
where $c_0$ is an integration constant,
which we express in terms of another constant $\sigma$ via
$
c_0={\rho_0}/{\sigma P_0(N+1)} \ ,
$
to obtain the more convenient expression
\begin{equation}
\Theta =\frac{\rho_0}{\sigma P_0(N+1)}(u -\sigma) \ .
\label{theta1}
\end{equation}
A convenient choice for the constant $\rho_0$ is given by $\rho_0=10^{-3}$.
The constant $P_0$ follows from Eq.~\ref{polEOSprho},
$P_0=K \rho_0^{\Gamma}$.

\subsection{Expansions, Boundary Conditions, Global Charges}

Having solved the equations for the neutron star matter in terms
of the metric functions, as given by Eq.~(\ref{theta1}), next
the PDEs for the metric functions and the dilaton function
need to be considered.
We therefore expand these equations at the origin
to obtain regularity conditions for the functions.
The expansion at the neutron star center reads
\begin{eqnarray}
\nu_i & = & \nu_{i c} + \nu_{i2}\frac{r^2}{2} + {\cal O}(r^3) , 
\ \ \ i=0,1,2 \ , 
\nonumber \\
\omega & = & \omega_{c} + \omega_{2}\frac{r^2}{2} + {\cal O}(r^3) , 
\nonumber \\
\phi & = & \phi_{c} + \phi_{2}\frac{r^2}{2} + {\cal O}(r^3) . 
\end{eqnarray}
We therefore require at the center the boundary conditions
\begin{equation}
\partial_r \left. \nu_i \right|_{r=0,\theta} = 0 \ , \ \ \ i=0,1,2 \ ,  \ \ \ 
\partial_r \left. \omega \right|_{r=0,\theta} = 0 \ ,  \ \ \ 
\partial_r \left. \phi \right|_{r=0,\theta} = 0 \ . 
\end{equation}

Note, that the central density and the central pressure
of the neutron star matter are determined by the integration
constant $\sigma$ in Eq.~(\ref{theta1}) and by the value
of metric function $\nu_0$ at the origin. 
At the surface of the neutron star the pressure
and thus the function $\Theta$ vanishes. 
However, we do not impose this outer boundary.
It follows from the integration. We only scan during the
integration procedure where $\Theta$ vanishes
(see the discussion below).

Since we are looking for asymptotically flat solutions,
we require, that the metric approaches the Minkowski metric
in the asymptotic region.
Here the metric functions and the dilaton function possess the expansion
\begin{eqnarray}
\nu_0 & = & -\frac{M}{2r} + \frac{D_1 M}{3 r^3} - \frac{M_2}{r^3} P_2(\cos\theta) +{\cal O}(r^{-4}) ,
\label{exnu0}\\
\nu_1 & = & \frac{D_1}{r^2} +{\cal O}(r^{-3}) ,
\label{exnu1}\\
\nu_2 & = & -\frac{4 M^2+16 D_1+ q^2}{8 r^2}\sin^2\theta +{\cal O}(r^{-3}) ,
\label{exnu2}\\
\omega & = & \frac{2 J}{r^3}  +{\cal O}(r^{-4}) ,
\label{exom}\\
\phi & = & \frac{q}{r}  +{\cal O}(r^{-2}) ,
\label{exdil}
\end{eqnarray}
where $P_2(\cos\theta)$ is the second Legendre polynomial.

From this expansion we can read off the global charges of the 
neutron star.
$M$ is the mass, $J$ is the angular momentum,
and $q$ is the dilaton charge.
The additional expansion constants $D_1$ and $M_2$ 
enter together with the mass $M$ and the dilaton
charge $q$ into the expression for 
the quadrupole moment $Q$ of the neutron star 
\cite{Kleihaus:2014lba}
\begin{equation}
Q 
=  -M_2 +\frac{4}{3}\left[\frac{1}{4}+\frac{D_1}{M^2}
+\frac{q^2}{16M^2}\right] M^3 .
\label{Q}
\end{equation}
(In Appendix A we give a brief derivation of the quadrupole moment.)

The boundary conditions in the asymptotic region follow from the expansion 
Eqs.~(\ref{exnu0})-(\ref{exdil}),
\begin{equation}
\nu_i \to 0 \ , \ \ \ i=0,1,2 \ ,  \ \ \ 
\omega \to 0 \ ,  \ \ \ 
\phi \to   0 \ . 
\end{equation}

Requiring regularity along the rotations axis ($\theta=0$, $\pi$) 
yields the boundary conditions
\begin{equation}
\partial_\theta \left. \nu_i \right|_{\theta=0 ,\pi} = 0 \ , \ \ \ i=0,1 \ ,  \ \ \ 
 \left. \nu_2 \right|_{\theta=0 ,\pi} = 0 \ ,  \ \ \ 
\partial_\theta \left. \omega \right|_{\theta=0 ,\pi} = 0 \ ,  \ \ \ 
\partial_\theta \left. \phi \right|_{\theta=0 ,\pi} = 0 \ .
\end{equation}
We also impose reflection symmetry with respect to the equatorial plane 
($\theta=\pi/2$).
The corresponding boundary conditions are given by
\begin{equation}
\partial_\theta \left. \nu_i \right|_{\theta=\frac{\pi}{2} } = 0 \ , \ \ \ i=0,1,2 \ ,  \ \ \ 
\partial_\theta \left. \omega \right|_{\theta=\frac{\pi}{2}  } = 0 \ ,  \ \ \ 
\partial_\theta \left. \phi \right|_{\theta=\frac{\pi}{2}  } = 0 \ .
\end{equation}

\subsection{Center and Surface of the Star}


The central pressure and the central energy density are obtained from
Eqs.~(\ref{polEOS}) together with Eqs.~(\ref{theta1}) and (\ref{u2}), 
evaluated at the center,
\begin{equation}
P_c = P_0 \Theta_c^{N+1}, \ \epsilon_c = N P_c + \rho_0 \Theta_c^N  , 
\ {\rm with} \
\Theta_c =\frac{\rho_0}{\sigma P_0(N+1)}(e^{-\nu_{0c}} -\sigma)  ,
\label{Peps_c}
\end{equation}
where the value of $\sigma$ is a free parameter.

The boundary of the star is defined as the surface 
where the pressure vanishes, or equivalently as noted above,
where the function $\Theta(r,\theta)$ vanishes. 
Let us parametrize the boundary by coordinates 
$(r_b(\theta),\theta,\vphi)$. 
Then the metric at the boundary (at fixed time) reads
\begin{equation}
ds_b^2 = 
       e^{2(\nu_{1b}-\nu_{0b})}\left(e^{2 \nu_{2b}}
       \left[(\partial_\theta r_b)^2+r_b^2\right] d\theta^2
       +r_b^2 \sin^2\theta d\varphi^2 \right) \  ,
\label{metricb}  
\end{equation}
where $\nu_{ib} = \nu_i(r_b(\theta),\theta)$. 

We use this metric to define the area $A_b$, 
the equatorial radius $R_e$ and the polar 
radius $R_p$ of the neutron star
\begin{eqnarray}
A_b & = & \int{\sqrt{\det{(g_b)}} } d\theta d\vphi = 
         4\pi \int_0^{\pi/2}r_b e^{2(\nu_{1b}-\nu_{0b})+\nu_{2b}}\sqrt{(\partial_\theta r_b)^2+r_b^2}
                    \sin\theta   d\theta \ , 
\label{area_b}\\
R_e & = & \frac{1}{2\pi}\int \sqrt{g_{b\vphi\vphi}} d\vphi = 
          \left(e^{(\nu_{1b}-\nu_{0b})}r_b\right)_{\theta=\pi/2} \ , 	    
\label{r_e}\\
R_p & = & \frac{1}{\pi}\int_0^{\pi}\sqrt{g_{b\theta\theta}} d\theta
        = \frac{2}{\pi}\int_0^{\pi/2}e^{\nu_{1b}-\nu_{0b}+\nu_{2b}}\sqrt{(\partial_\theta r_b)^2+r_b^2} 
	d\theta \ . 
\label{r_p} 
\end{eqnarray}

Let us finally address the Kepler limit of a neutron star. 
This limit is reached when the neutron star rotates so rapidly,
that the angular velocity $\Omega$ of the fluid reaches
the angular velocity $\Omega_p$ 
of a free particle on the equator of the boundary.
Then a fluid element on the surface of the star at the equator
is no longer bound and the star starts to dissolve.
Hence the Kepler limit is also called mass-shedding limit 
and forms a part of the boundary of the physically relevant domain
of neutron stars.

The Kepler angular velocity $\Omega_K$ is found by considering 
the circular orbit of a massive particle at the equator of the star
\cite{Friedman:2013xza}.
The geodesic equation yields for the angular velocity 
$\Omega_p$ of the particle
\begin{equation}
(\Omega_p-\omega_b)^2 - 2 a_p (\Omega_p-\omega_b) + b_p = 0 \ ,
\end{equation}
where $\omega_b =\omega(r_b,\pi/2)$ and $a_p$, $b_p$ are expressions 
in the metric functions and their derivatives at $(r_b,\pi/2)$,
\begin{equation}
a_p = 
\left. 
\frac{r\partial_r \omega}{2\left(1-r\partial_r(\nu_0-\nu_1)\right)} 
\right|_{r_b,\pi/2}
\ , \ \ \ 
b_p = 
\left. 
-\frac{e^{4\nu_0-2\nu_1}r\partial_r \nu_0}{r^2\left(1-r\partial_r(\nu_0-\nu_1)\right)}
\right|_{r_b,\pi/2} \ .
\end{equation}
Solving for $\Omega_p$ one finds
\begin{equation}
\Omega_p = \omega_b + a_p +\sqrt{a_p^2 - b_p} \ .
\label{OmKep}
\end{equation}
The Kepler angular velocity $\Omega_K$ is reached for
$\Omega_K= \Omega = \Omega_p$,
since for $\Omega > \Omega_p$ the star would start losing mass.

Note, that the rotational period $T$ of the neutron star is related to the 
angular velocity $\Omega$ by $ T[s] = 0.03952/\Omega$ 
for our choice of units.

\section{Neutron Stars in Einstein-Gauss-Bonnet-Dilaton Theory}

Let us now turn to the presentation of our numerical results.
Note, throughout this work we have set the dilaton coupling parameter 
$\gamma$ to one, thus employing the value from heterotic string theory
(in our conventions). The dependence on $\gamma$ was studied 
previously in \cite{Pani:2011xm} for static and slowly rotating
neutron stars, showing that it is basically only the product $\alpha \gamma$
of the two coupling constants, which determines the neutron star properties.

For a fixed EOS and a fixed value of the GB coupling $\alpha$,
the solutions for the rotating neutron stars then depend on two parameters.
One parameter is the angular velocity of the neutron star fluid $\Omega$,
while the second parameter determines the central energy density $\epsilon_c$
and the central pressure $P_c$ of the the neutron star fluid.
As seen from Eqs.~(\ref{Peps_c}), 
both depend on the integration constant $\sigma$, which we therefore
employ as the second parameter to be varied in the calculations.
Although the physical meaning of $\sigma$ is not obvious,
Eq.~(\ref{theta1}) shows that $\sigma$ is the time component of the
four velocity of the fluid at the boundary of the star,
where $\Theta$ vanishes. 

Concerning the GB coupling constant $\alpha$
we here consider the following three values
for the dimensionless quantity as defined in (\ref{ab}): 
$\alpha=0$ (GR limit), $\alpha=1$ and $\alpha=2$.
Thus we stay below the upper bound obtained 
from low mass x-ray binaries \cite{Yagi:2012gp},
which would correspond to $\alpha = 12$ when
converted to our dimensionless $\alpha$.
(Note, that in \cite{Yagi:2012gp} the bound is given in units of length as
$\sqrt{|\alpha_Y|}=1.9 \times 10^5$ cm.)
The (EOS dependent) bound extracted in \cite{Pani:2011xm}, as obtained 
by requiring the existence of sequences of static neutron stars solutions
up to a maximum mass, on the other hand, would 
correspond to $\alpha = 3.36$.
(In the dimensionful units of \cite{Pani:2011xm}
it is given as $\alpha_P = 23.8 M_\odot^2$.)
Note that the constraints from the solar system are
much weaker \cite{Pani:2009wy,Yagi:2012gp,Yagi:2015oca}.

Since the calculations for the rapidly rotating neutron stars
and, in particular, the extraction of the Kepler limit
have proven to be very time consuming, we have considered
only two EOSs here, FPS \cite{Lorenz:1992zz,Haensel:2004nu} and DI-II \cite{Diaz-Alonso:1985},
as discussed below.
Moreover, we have restricted the calculations
to solutions with masses $M\geq M_\odot$ for the EOS FPS
and $M\geq 1.4 M_\odot$ for the EOS DI-II. 

In the following we briefly address the numerical method and
the EOS. We then present our results for the mass-radius relation and
the mass-central density relation. 
In particular we exhibit the physically relevant domain
for rotating neutron stars.
Subsequently, we consider the compactness, the angular momentum,
the rotation period and the dilaton charge.
We then extract the moment of inertia and the quadrupole moment,
and consider their universal relation.
Finally, we address the shape and deformation of the rapidly 
rotation neutron stars.

\subsection{Numerical Method}

Let us now turn to the numerical method employed in the construction
of the rapidly rotating neutron star solutions.
Before starting the numerical procedure 
we introduce the compactified coordinate $x$ via
\begin{equation}
 r = \hat{r}_0\frac{x}{1-x}  , \ 0\leq x \leq 1  ,
\end{equation}
thus mapping the (semi)infinite interval of the coordinate $r$ 
to the unit interval.
A convenient choice for the constant is $\hat{r}_0 = 10$.
Then the domain of integration $[0,1]\times [0,\pi/2]$ 
is subdivided into $N_x$ subintervals
is $x$ direction and $N_\theta$ subintervals in $\theta$ direction. 
Typical mesh sizes are $N_x = 260$, $N_\theta=60$.

We then employ the CADSOL package \cite{schoen}
based on the Newton-Raphson method. 
The partial derivatives are discretized with sixth order of consistency.
We need to apply a special treatment at the boundary of the star.
When evaluating the PDEs and their Jacobian 
we check for each $\theta$ and $x$, 
whether the function $\Theta$ is positive or negative. 
If $\Theta<0$ the meshpoint is outside the star. 
In this case the parameter $\beta$ (\ref{ab}) is set to zero. 
Otherwise $\beta$ is set to one.

In the static limit the Einstein and field equations 
reduce to ordinary differential equations. 
In this case we mainly use the COLSYS package \cite{COLSYS} 
to compute the neutron star solutions. 
For vanishing $\alpha$ we obtain neutron stars in GR. 
Here we employ the rns code \cite{Friedman:2013xza}, 
except for large values of the central density,
where the rns code does not converge. 
Employing these different methods allows us 
to compare with the respective results obtained with CADSOL.
In all cases we find excellent agreement.

\subsection{Equations of State}

For the rapidly rotating neutron stars 
in dEGB theory we here consider two EOSs.
Both have the advantage of being relatively simple,
while they have the disadvantage that the maximum mass of their
static sequence is below $2 M_\odot$. Thus they cannot
describe (slowly rotating) high mass neutron stars 
\cite{Demorest:2010bx,Antoniadis:2013pzd}.

\begin{figure}[h!]
\begin{center}
\includegraphics[height=.25\textheight, angle =0]{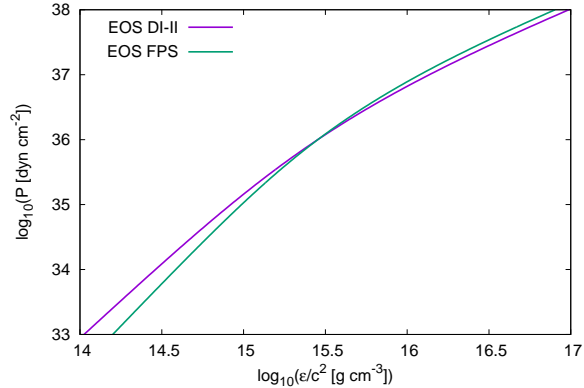}
\end{center}
\vspace{-0.5cm}
\caption{
The pressure-energy density relation for the two EOSs, FPS and DI-II, employed
for the neutron stars.
The pressure $P$ is given in units of dyn/cm$^2$, the
energy density $\epsilon/c^2$ in units of g/cm$^3$.
\label{fig0}
}
\end{figure}

The first EOS corresponds to a polytropic EOS,
as described by Eq.~(\ref{polEOSprho}),
with $N=0.7463$ and $K=1186.0$ 
(with our choice of units). 
This EOS is denoted by DI-II and taken from \cite{Diaz-Alonso:1985}.
It has been widely used in neutron star physics in GR 
as well as in scalar-tensor theory
\cite{Damour:1993hw,Doneva:2013qva}.

The second EOS represents an approximation
to the FPS EOS from \cite{Lorenz:1992zz},
where the analytical fit \cite{Haensel:2004nu} to the FPS EOS 
is approximated by a fit to a polytropic EOS
with $N=0.6104$ and $K=5392.0$
\cite{Haensel:2004nu}.
Note, that in \cite{Blazquez-Salcedo:2015ets} a set of eight
realistic EOSs was employed to obtain sequences 
of static neutron stars in dEGB theory
and to study the effect of the dilaton and the GB term.

\subsection{Physical Domain of Neutron Star Solutions}

\begin{figure}[h!]
\begin{center}
\mbox{
(a)\includegraphics[height=.25\textheight, angle =0]{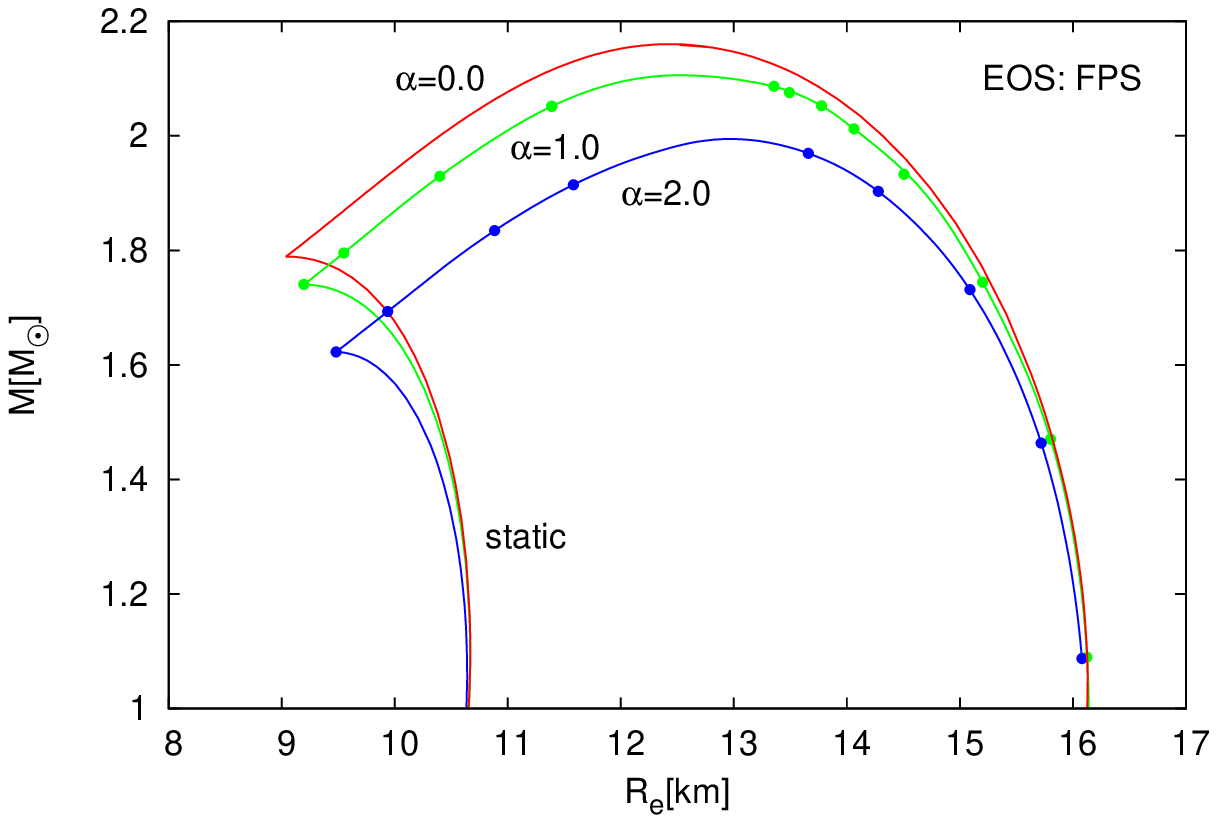}
(b)\includegraphics[height=.25\textheight, angle =0]{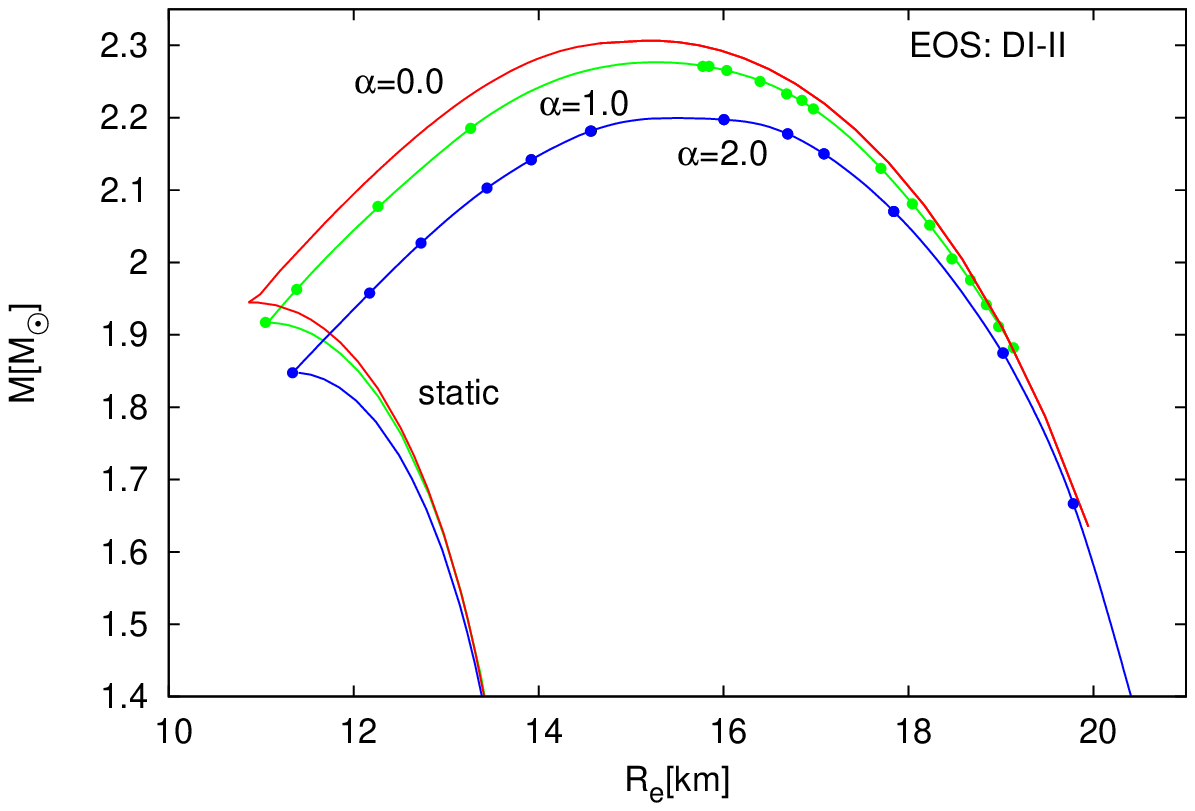}
}
\end{center}
\vspace{-0.5cm}
\caption{
(a) The physically relevant domain
is shown for the mass-radius relation for $\alpha=0$, 1 and 2
for the EOS FPS.
For a given $\alpha$ the left boundary curve represents the sequence of static solutions,
while the right boundary curve represents the sequence of neutron stars rotating at the
Kepler limit. Both are connected by the secular instability line.
The mass $M$ is given in units of the solar mass $M_\odot$ and the
equatorial radius $R_e$ in units of kilometers.
(b) Same as (a) for the  EOS DI-II.
\label{fig1}
}
\end{figure}

Let us now address the sequences of neutron star solution,
which delimit the physically relevant domain.
This domain is exhibited for the mass-radius relation in Fig.~\ref{fig1}
for the two EOSs employed, the EOS FPS and the EOS DI-II, respectively,
and the values of the dimensionless GB coupling constant
$\alpha=0$, 1 and 2.
This domain is delimited by 
(i) the set of neutron stars rotating at the Kepler limit,
(ii) the set of neutron stars along the secular instability line,
formed by the set of neutron stars with maximum mass at fixed angular momentum,
and (iii) the set of static neutron stars.
In the following we address these limiting curves in more detail.

For a given $\alpha$ the left boundary curve of this domain for the mass-radius relation
represents the sequence of static solutions ($\Omega=0$), 
where the mass increases monotonically
with decreasing radius until the maximal value of the mass of a static neutron star
is reached. The FPS EOS is a rather soft EOS, therefore this maximal value of the mass is
rather low already in GR. As the GB term is coupled, the mass is decreased
monotonically with increasing GB coupling $\alpha$.
When continuing beyond the maximum mass the set of neutron stars would no longer
be stable but exhibit a first radially unstable mode.
These static solutions are spherically symmetric, thus the equatorial radius $R_e$
agrees with the polar radius $R_p$.

As shown in \cite{Pani:2011xm} $\alpha$ cannot increase arbitrarily,
while still giving a complete sequence of static neutron stars.
Instead, beyond a critical value of $\alpha$ the sequence no longer reaches
a maximum but ends in a critical configuration, when a certain
radicand in the expansion of the dilaton field vanishes.
(Note, that the occurrence of such critical values was noted first for black holes
in dEGB theory \cite{Kanti:1995vq}.)
This observation has been used in \cite{Pani:2011xm} to obtain an in principle
EOS dependent bound for $\alpha$.
The maximal value of $\alpha$ chosen here is still below this bound
(see the discussion above).

The right boundary curve of this domain for the mass-radius relation
represents the sequence of neutron stars rotating at the Kepler limit
 ($\Omega=\Omega_K$). 
When the neutron star is rotating at the Kepler limit,
its fluid elements at the neutron star boundary at the equator are no longer
bound. At a slighly faster rotation rate they would be shed,
and the neutron star would no longer be stable.
In order to obtain the Kepler limit with high precision we compute sequences of solutions
for a fixed parameter $\sigma$ and increasing values of $\Omega$, 
while monitoring the quantity $\delta = 1-\Omega_p/\Omega$. 
For small $\delta$ we then consider the physical quantities
mass, angular momentum, equatorial radius, etc as functions of $\delta$ 
and extrapolate to $\delta=0$.

Both boundary curves are connected by the secular instability line,
which forms the remaining upper part of the boundary
of the physically relevant domain,
and extends from the maximum of the static sequence
to the Kepler sequence.
Here, analogous to the static sequence, the neutron stars
become unstable at the maximal value of the mass for a 
fixed value of the angular momentum
\cite{Friedman:1988er}.

In Fig.~\ref{fig1}, where the mass-radius relation is shown for these
three boundary curves, 
the dots represent the calculated values for rotating dEGB neutron stars
with maximum mass along the secular instability line and at the Kepler limit.
The solid curves for $\alpha=1$ and 2 interpolate between
these points and also include the static sequence.
We recall, that for $\alpha=0$ we used the more efficient rns code.

The mass-radius dependence on the boundary is then as follows:
(i) For the static neutron stars the mass increases with decreasing radius 
up to the stability limit.  
(ii) Along the secular instability line the mass increases 
with increasing radius
until (at the global maximum of the mass
in this domain) the Kepler limit is reached.
(iii) For the neutron stars at the Kepler limit, the mass then decreases 
with increasing radius. 

This qualitative behaviour is common to neutron stars in GR
and in dEGB theory.
We observe as a general feature of the dEGB neutron stars that their
physically relevant domain decreases
as the GB coupling $\alpha$ increases.
Thus, the maximum masses are smaller for larger values of $\alpha$ while 
the minimum radii are larger. 
For small masses and large radii the Kepler limit
is (almost) independent of $\alpha$ (as long as it exists). 

Comparing these domains for the two EOSs we conclude
that analogous to GR also for dEGB theory
neutron stars are
larger and more massive for EOS DI-II 
than for EOS FPS.

\subsection{Mass-Radius Relation and Mass-Energy Density Relation}

\begin{figure}[h!]
\begin{center}
\mbox{
(a)\includegraphics[height=.25\textheight, angle =0]{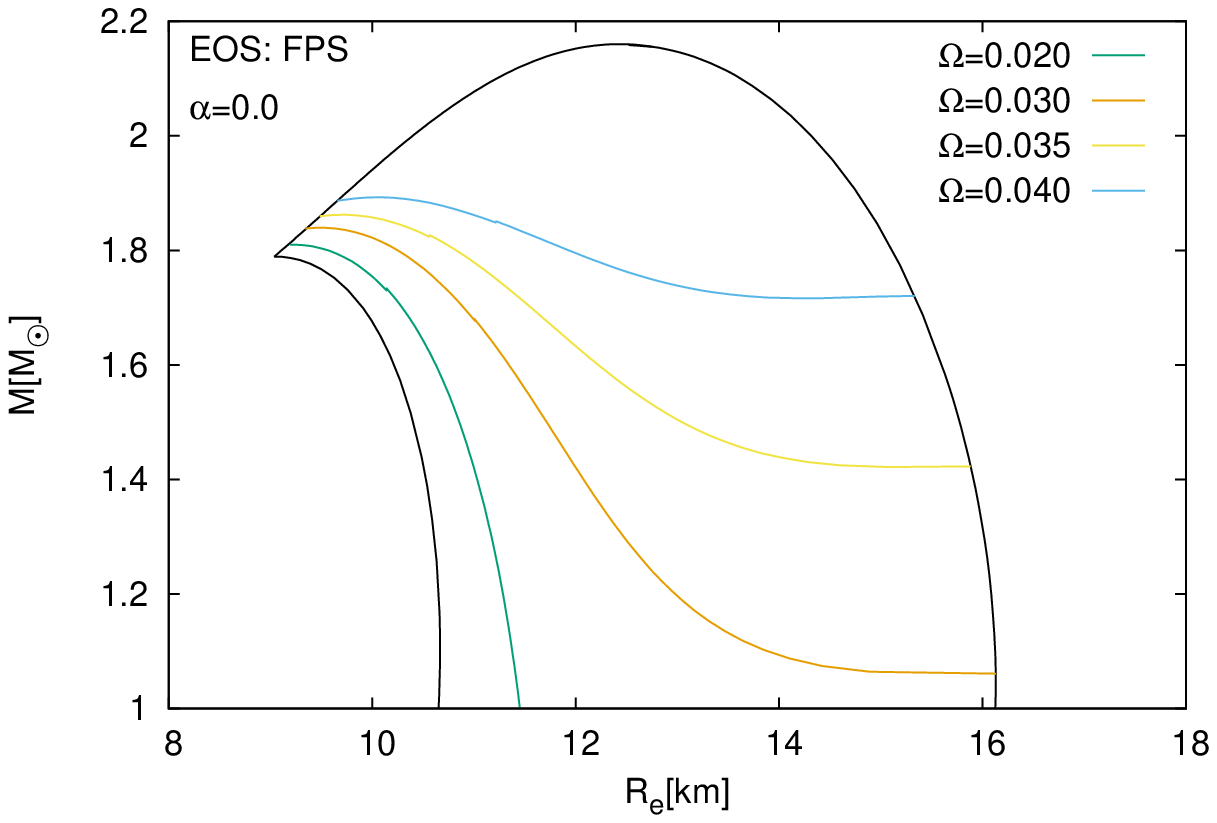}
(d)\includegraphics[height=.25\textheight, angle =0]{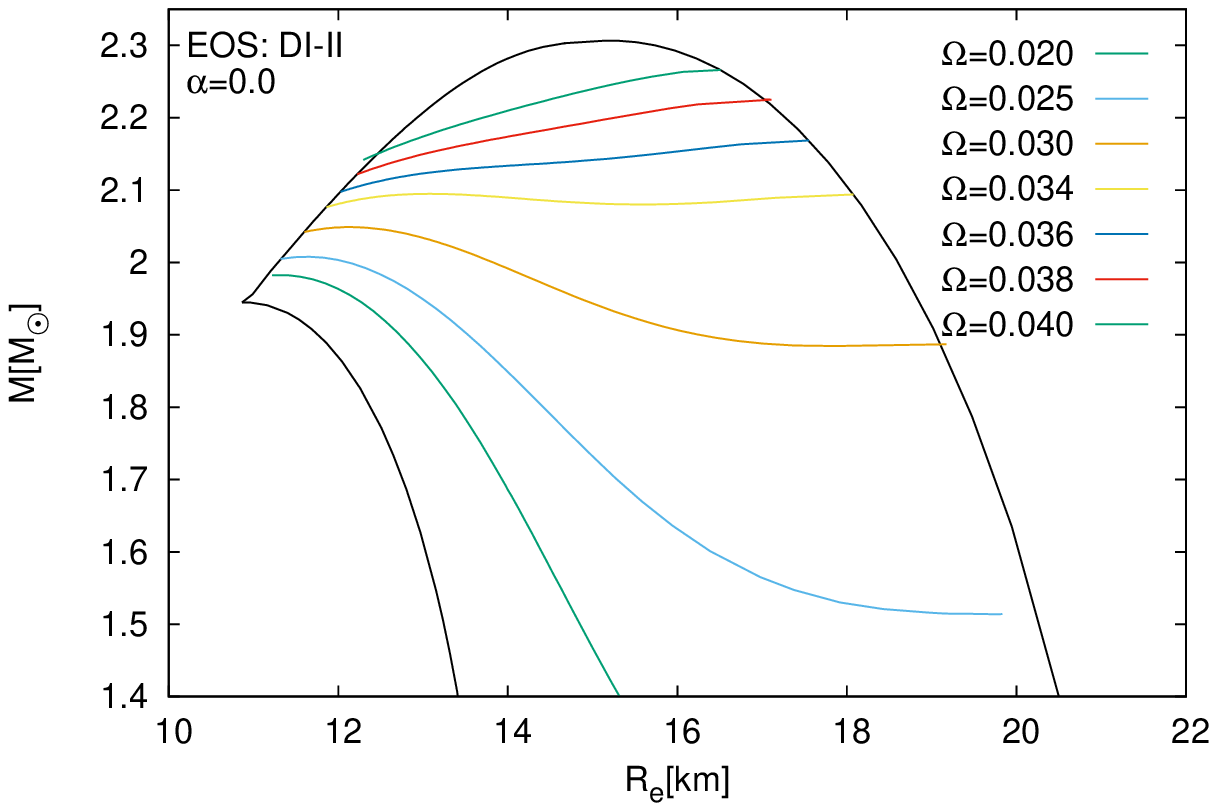}
}
\mbox{
(b)\includegraphics[height=.25\textheight, angle =0]{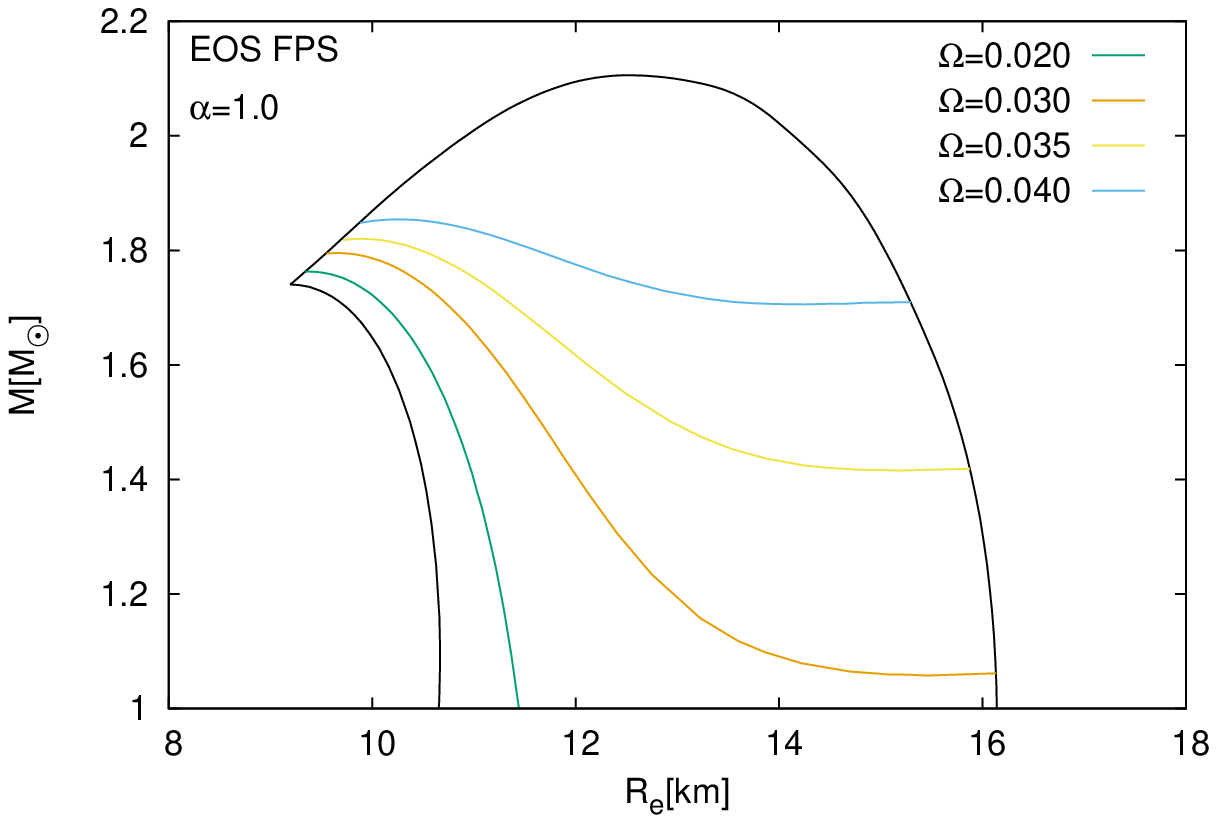}
(e)\includegraphics[height=.25\textheight, angle =0]{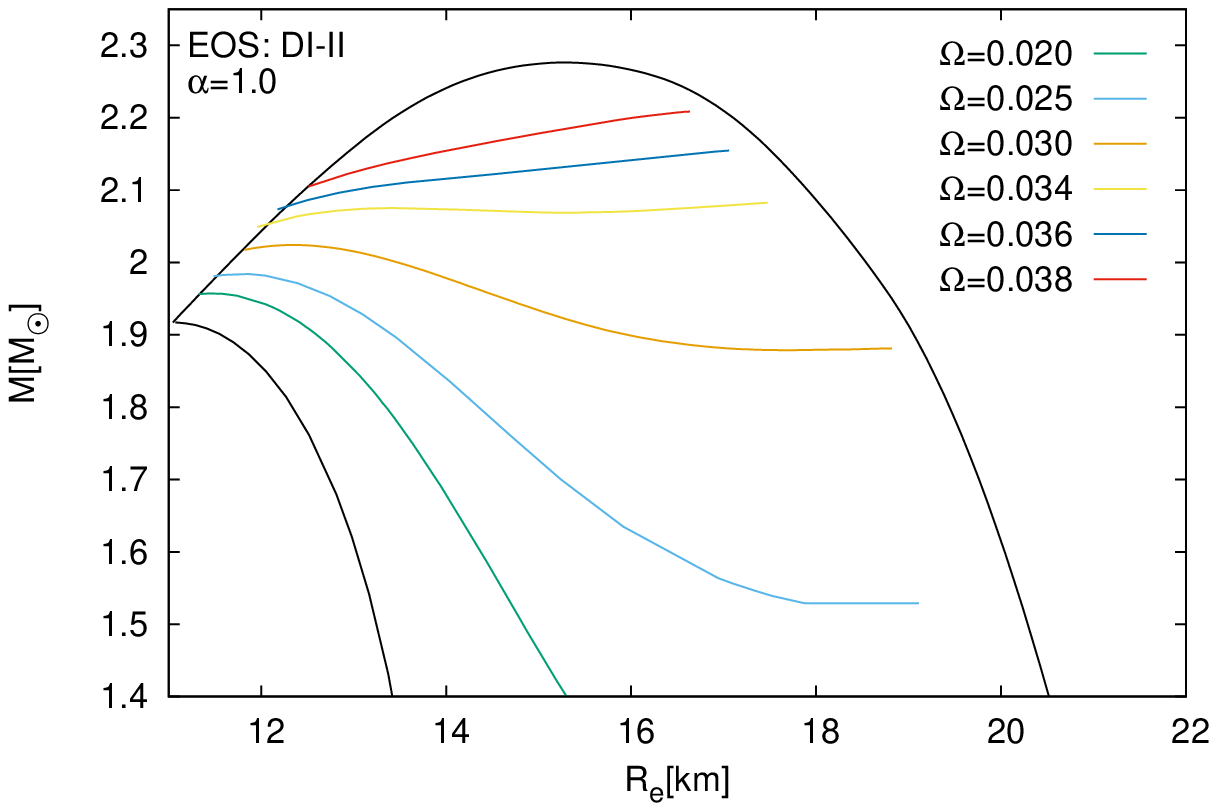}
}
\mbox{
(c)\includegraphics[height=.25\textheight, angle =0]{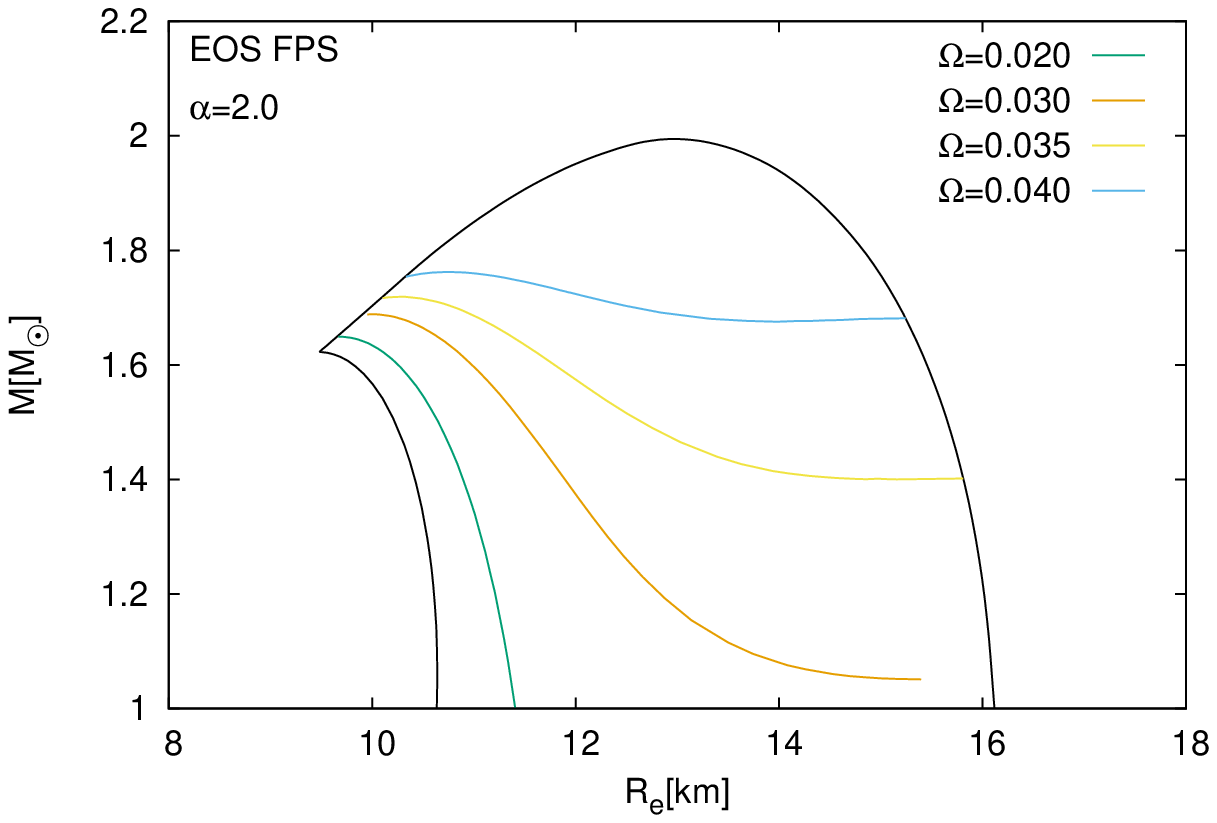}
(f)\includegraphics[height=.25\textheight, angle =0]{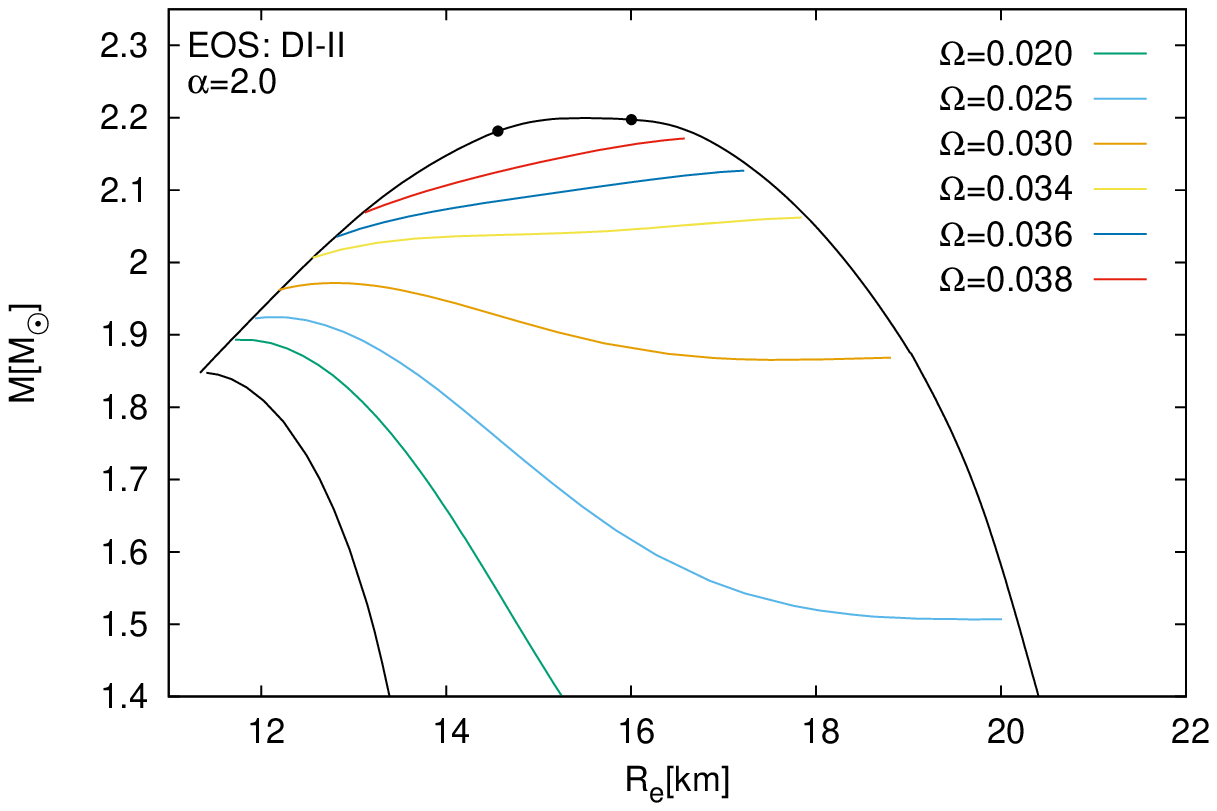}
}

\end{center}
\vspace{-0.5cm}
\caption{
(a-c) The mass-radius relation of neutron stars in the physically relevant domain
for several values of the dimensionless angular velocity $\Omega$ 
for the EOS FPS. ($\Omega=0.01$ corresponds to $f =323$ Hz.)
The mass $M$ is given in units of the solar mass $M_\odot$ and the
equatorial radius $R_e$ in units of kilometers.
The GB coupling constant $\alpha$ has the values $\alpha=0$, 1 and 2.
The solid black line represents the sequence of static neutron stars,
the secular instability line and the sequence of neutron stars at the
Kepler limit. 
(d-f) Same as (a-c) for the EOS DI-II. 
\label{fig2}
}
\end{figure}

Having determined the limits where the secular instability and the mass shedding
set in, we now  discuss the mass-radius relation in more detail. 
To this end we exhibit in Fig.~\ref{fig2}, 
the mass-radius relation of sequences of neutron stars with 
fixed angular velocity $\Omega$. 
We note, that the values of $\Omega$ in the figure are given in dimensionless units.
$\Omega=0.01$ there corresponds to a frequency of $f =323$ Hz.
For comparison we recall that the fastest rotating pulsar has a frequency of
$\nu=716$ Hz
\cite{Hessels:2006ze}.

The mass-radius relation of rotating neutron stars in GR has
been recently readdressed in \cite{Cipolletta:2015nga},
where besides the static and the Keplerian sequence
also sequences of neutron stars rotating at fixed angular velocity
have been constructed numerically,
varying the frequency from $f=50$ Hz to $f=716$ Hz
for several EOSs.
While the $f=50$ Hz sequence basically agrees with the static sequence,
small deviations start to arise as the frequency is increased,
and the equatorial radius increases slightly with increasing frequency.
This is expected since the rotation then starts to deform the star.
At $f=200$ Hz the deviation from the static sequence is still small,
but for larger values of the frequency the effect of rotation 
changes the mass-radius relation considerably \cite{Cipolletta:2015nga}.
For all values of $f$ considered in \cite{Cipolletta:2015nga} the mass decreases
with increasing radius.

We here do not address the small frequencies, where the mass-radius relation
hardly deviates from the static sequence. Instead, we consider sequences of
rapidly rotating neutron stars. We observe in Fig.~\ref{fig2} the same
monotonically decreasing behavior of the mass versus the 
equatorial radius for even larger values of the frequency. 
In contrast,
for very large frequencies the mass increases with the radius.
Fig.~\ref{fig2} shows that
the mass-radius sequences for fixed $\Omega$ show an analogous 
behavior independent of the value of the GB coupling employed
and independent of the EOS.

\begin{figure}[h!]
\begin{center}
\mbox{
(a)\includegraphics[height=.25\textheight, angle =0]{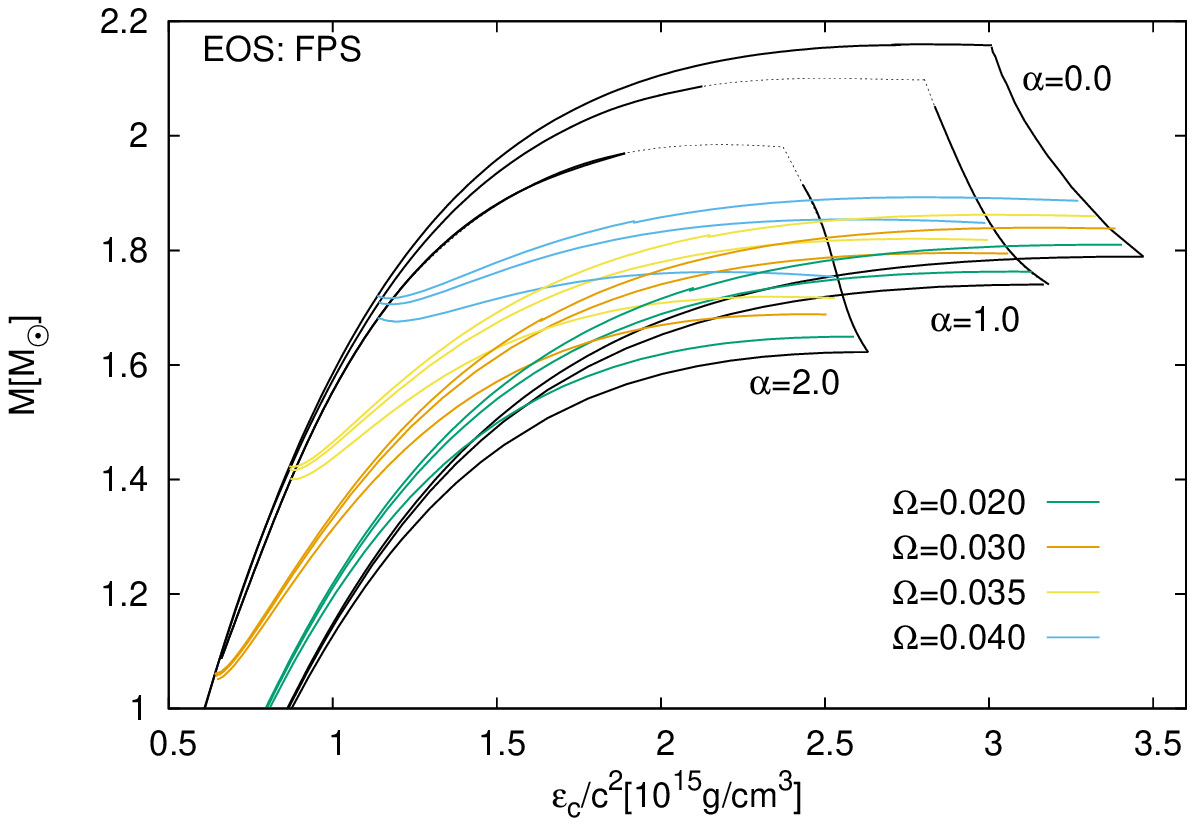}
(b)\includegraphics[height=.25\textheight, angle =0]{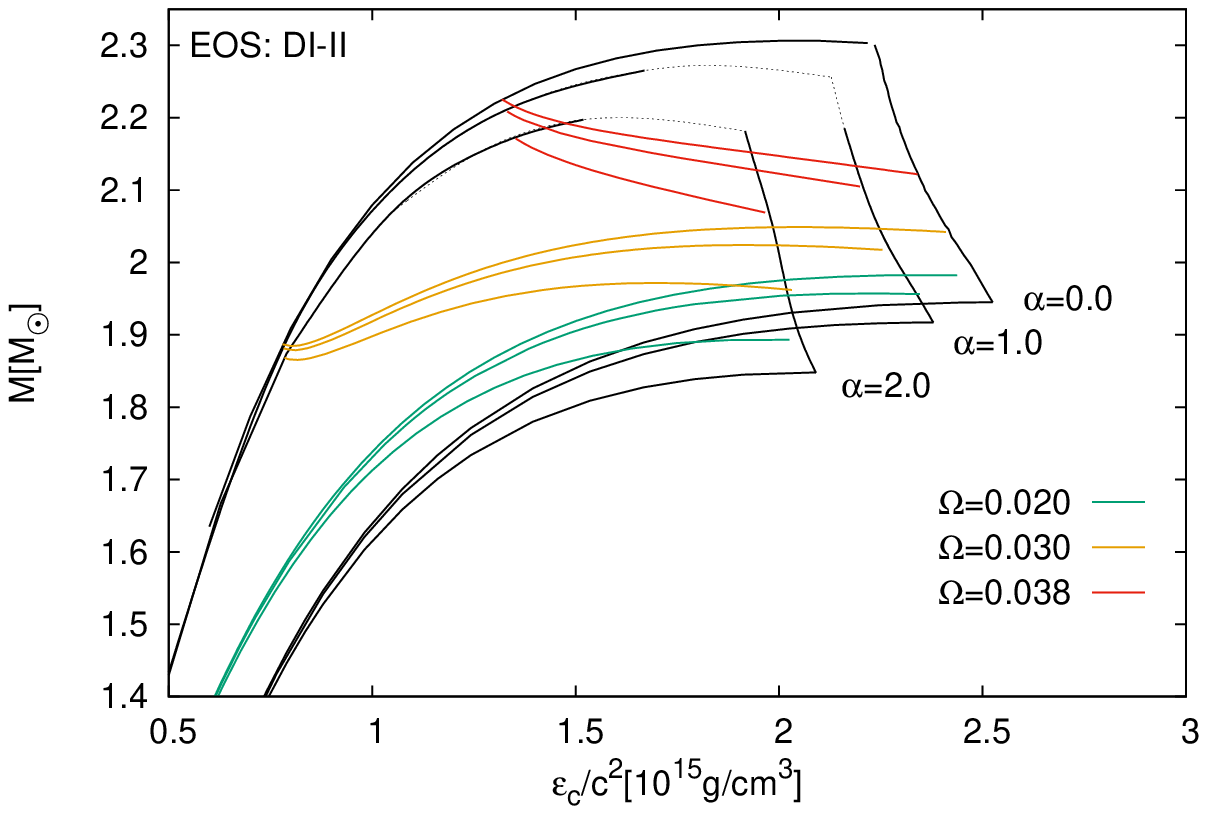}
}
\end{center}
\vspace{-0.5cm}
\caption{
(a) The mass-central energy density relation of neutron stars 
in the physically relevant domain
for several values of the dimensionless angular velocity $\Omega$ 
for the EOS FPS. ($\Omega=0.01$ corresponds to $f =323$ Hz.)
The mass $M$ is given in units of the solar mass $M_\odot$ and the
central value of the energy density $\epsilon_c/c^2$ in units of $10^{15}$ g/cm$^3$.
The GB coupling constant $\alpha$ has the values $\alpha=0$, 1 and 2.
The solid black lines represent the sequences of static neutron stars,
the secular instability lines and the sequences of neutron stars at the
Kepler limit. The dotted curves represent extrapolations indicating the expected
behaviour.
(b) Same as (a) for the  EOS DI-II. 
\label{fig8}
}
\end{figure}

As an alternative representation of our results
we exhibit in Fig.~\ref{fig8} the mass $M$ versus the central energy density 
$\epsilon_c$.
For a better comparison and extraction of the influence
of the GB coupling, we here include all sequences for a given EOS and all 
considered values of the
GB coupling $\alpha=0$, 1 and 2 in a single plot. 
The central energy density 
is maximal for static neutron stars in the stability limit
and decreases with increasing angular velocity
along the secular instability line.
Concerning the $\alpha$ dependence we note that
the maximum of the central energy density decreases with increasing $\alpha$. 
Close to the Kepler limit the mass-central energy density
relation is almost independent of $\alpha$. 
Comparison of the two EOSs shows that 
the EOS FPS yields neutron stars with larger central energy density than the EOS DI-II.

\subsection{Compactness, Angular Momentum, Rotation Period and Dilaton Charge}

The compactness of neutron stars is another quantity of considerable physical interest.
Let us define the compactness $C$ of rotating neutron stars as the 
ratio of the mass $M$ to the equatorial radius $R_e$ 
\begin{equation}
C = \frac{2GM}{R_e c^2} ,
\label{comp}
\end{equation}
normalized such that for black holes with mass $M$ and horizon radius $R_e$ 
the compactness would correspond to $C=1$.
(Note, that 
Kerr black holes have $C=1$ independent of the angular velocity of the horizon.)
With this normalization the compactness of neutron stars as discussed in \cite{Rhoades:1974fn}
is bounded from above by $C_{\rm lim} = 0.6706$.

\begin{figure}[h!]
\begin{center}
\mbox{
(a)\includegraphics[height=.25\textheight, angle =0]{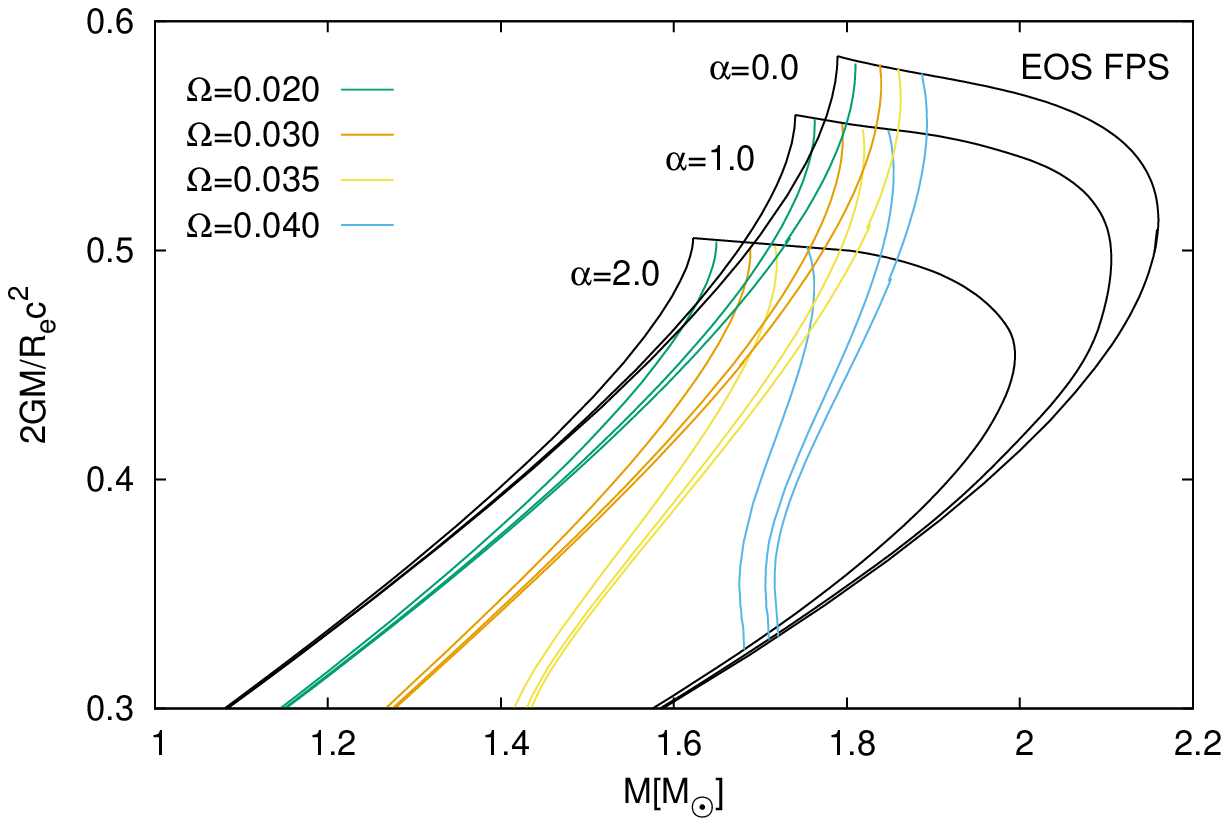}
(d)\includegraphics[height=.25\textheight, angle =0]{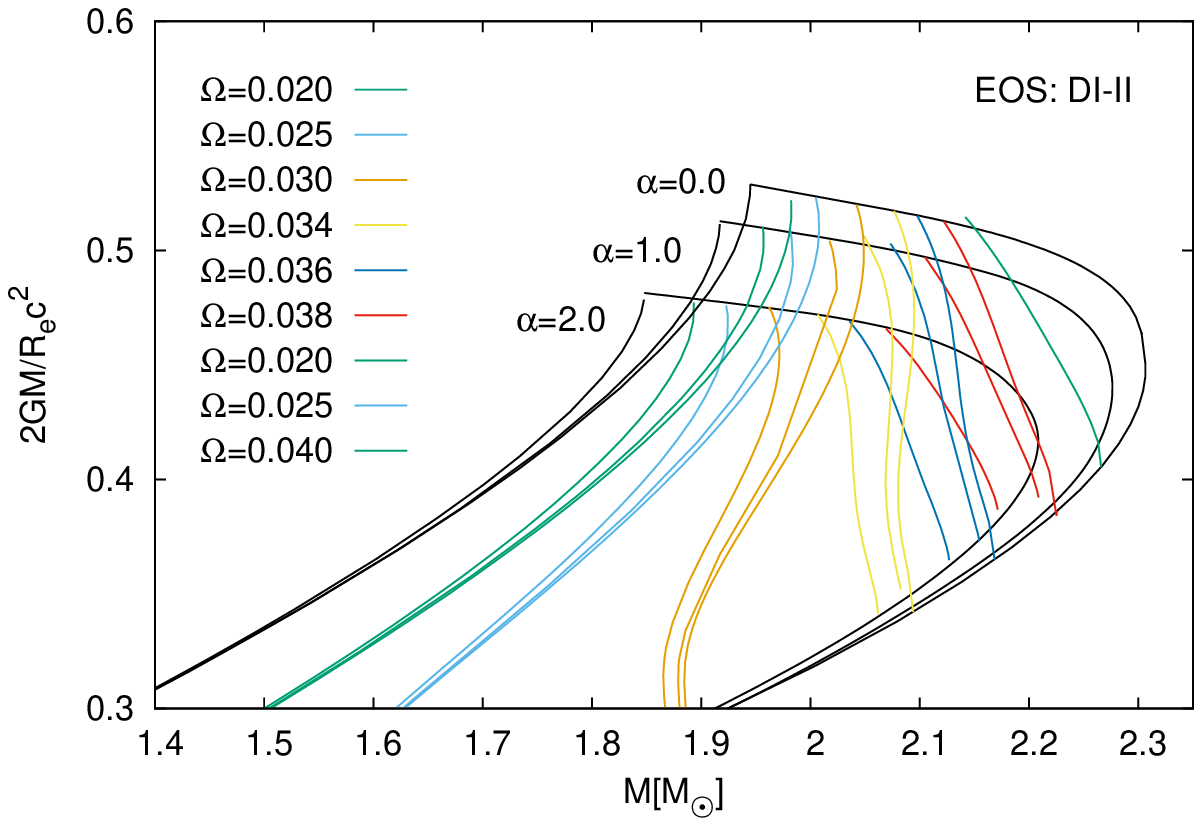}
}
\end{center}
\vspace{-0.5cm}
\caption{
(a) The compactness $C$, Eq.~(\ref{comp}), of neutron stars 
versus the mass in the physically relevant domain
for several values of the dimensionless angular velocity $\Omega$ 
for the EOS FPS. ($\Omega=0.01$ corresponds to $f =323$ Hz.)
The mass $M$ is given in units of the solar mass $M_\odot$.
The GB coupling constant $\alpha$ has the values $\alpha=0$, 1 and 2.
The solid black lines represent the static sequences,
the secular instability lines and the Keplerian sequences. 
(b) Same as (a) for the  EOS DI-II.
\label{fig3}
}
\end{figure}

We exhibit the compactness $C$ as a function of the mass in Fig.~\ref{fig3}
for several sequences of neutron stars 
with fixed angular velocity $\Omega$. 
The most compact neutrons stars are the static ones
with maximum mass. As the stars rotate, they can get more massive,
but at the same time their equatorial radius increases more rapidly,
so that the compactness decreases.
We note, that for all solutions the compactness is well below the limit $C_{\rm lim}$.
Considering the dependence on the GB coupling constant $\alpha$, we observe that
the compactness of the neutron stars decreases with increasing $\alpha$.
Generically, the EOS DI-II yields less compact neutron stars than the EOS FPS.

\begin{figure}[h!]
\begin{center}
\mbox{
(a)\includegraphics[height=.25\textheight, angle =0]{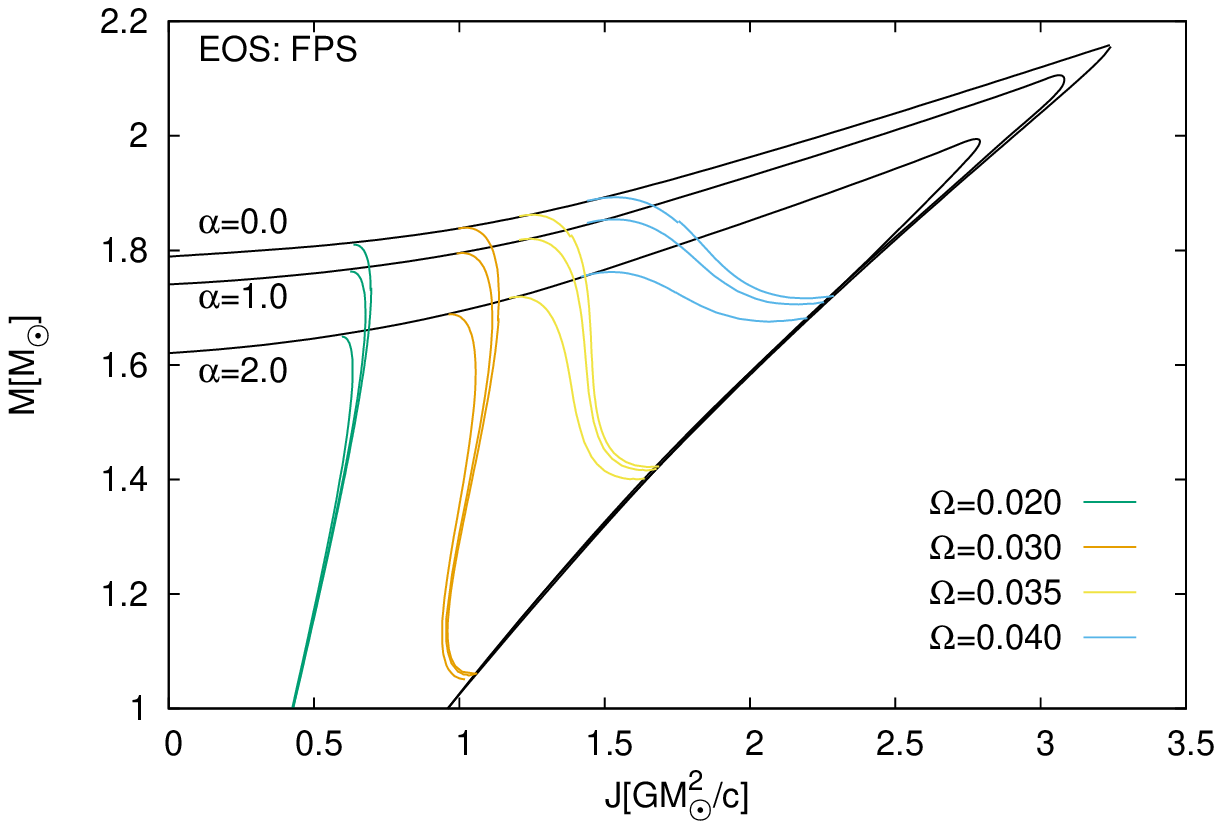}
(b)\includegraphics[height=.25\textheight, angle =0]{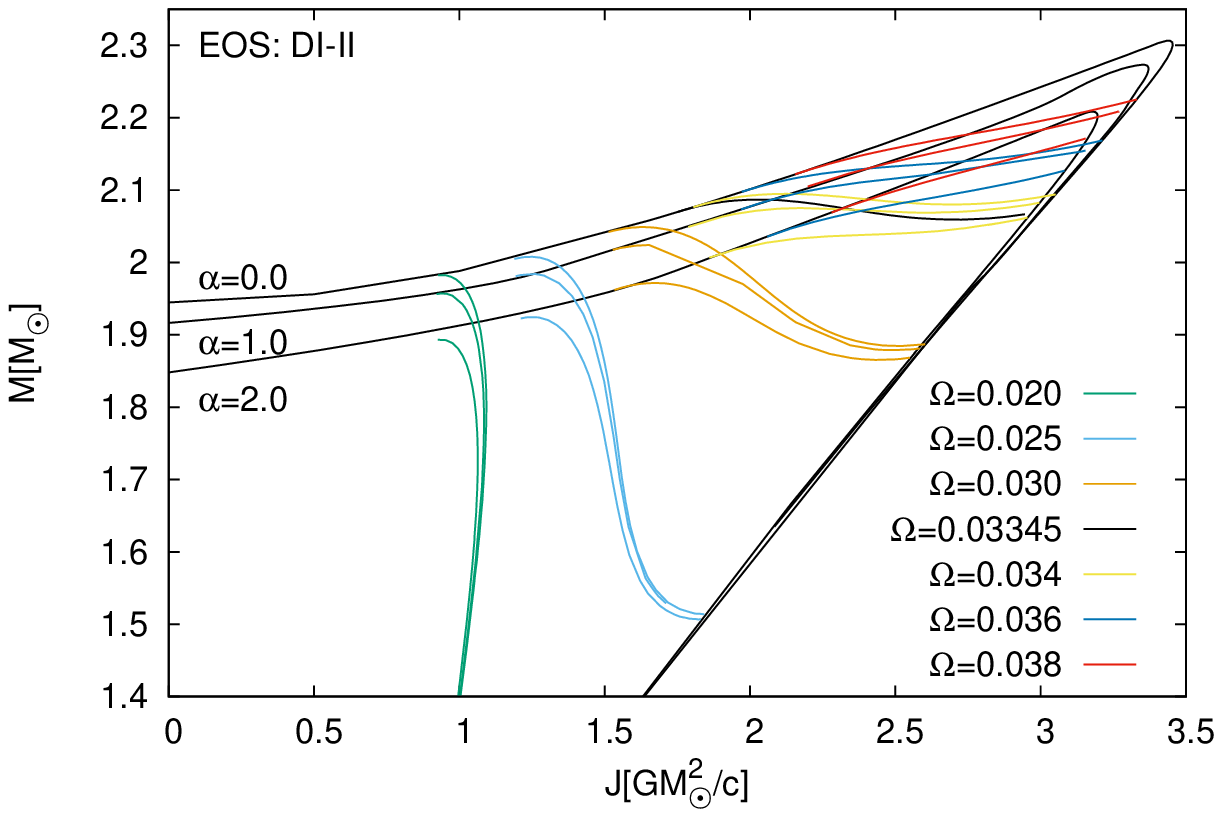}
}
\end{center}
\vspace{-0.5cm}
\caption{
(a) The mass $M$ is shown versus the angular momentum $J$ for neutron stars 
in the physically relevant domain
for several values of the dimensionless angular velocity $\Omega$ 
for the EOS FPS. ($\Omega=0.01$ corresponds to $f =323$ Hz.)
The mass $M$ is given in units of the solar mass $M_\odot$
the angular momentum $J$ in unist of $G M^2_\odot/c$.
The GB coupling constant $\alpha$ has the values $\alpha=0$, 1 and 2.
The solid black lines represent the static sequences,
the secular instability lines and the Keplerian sequences. 
(b) Same as (a) for the  EOS DI-II.
\label{fig4}
}
\end{figure}

Let us next address the
angular momentum $J$ of rotating neutron stars.
We exhibit in Fig.~\ref{fig4} the mass $M$
in units of the solar mass $M_\odot$ versus the angular momentum
in units of $G M^2_\odot/c$
for the set of GB coupling constants $\alpha=0$, 1 and 2.
Here the secular instability lines form the upper limit for the mass
while the lower limit is given by the Keplerian sequences.
The figures reveal an almost linear relation between the mass and the
angular momentum for the Keplerian sequences,
which is basically independent of $\alpha$.

Let us now compare with the angular momentum of black holes.
For Kerr black holes the reduced dimensionless angular momentum  $a/M$
\begin{equation}
\frac{a}{M}=\frac{cJ}{GM^2}
\label{aj}
\end{equation}
is limited by its value at maximum rotation $a_{\rm max}/M = 1$,
attained only by extremal black holes.
Interestingly, for dEGB black holes, this Kerr bound can be slightly exceeded
\cite{Kleihaus:2011tg,Kleihaus:2015aje}.

In \cite{Cipolletta:2015nga} the reduced dimensionless angular momentum
$a/M$ has been extracted for the Keplerian sequence of eleven EOSs,
including the EOS FPS.
For all EOSs considered the maximal value reached for
$a/M$ is about 0.7, i.e., distinctly below the Kerr value.
Moreover, except for large values of the mass,
the reduced dimensionless angular momentum $a/M$
varies only little with the mass.
For the EOS employed here, we observe a somewhat
different behavior, which may arise from their polytropic character
\cite{Cook:1993qj,Cook:1993qr}.

\begin{figure}[h!]
\begin{center}
\mbox{
(a)\includegraphics[height=.25\textheight, angle =0]{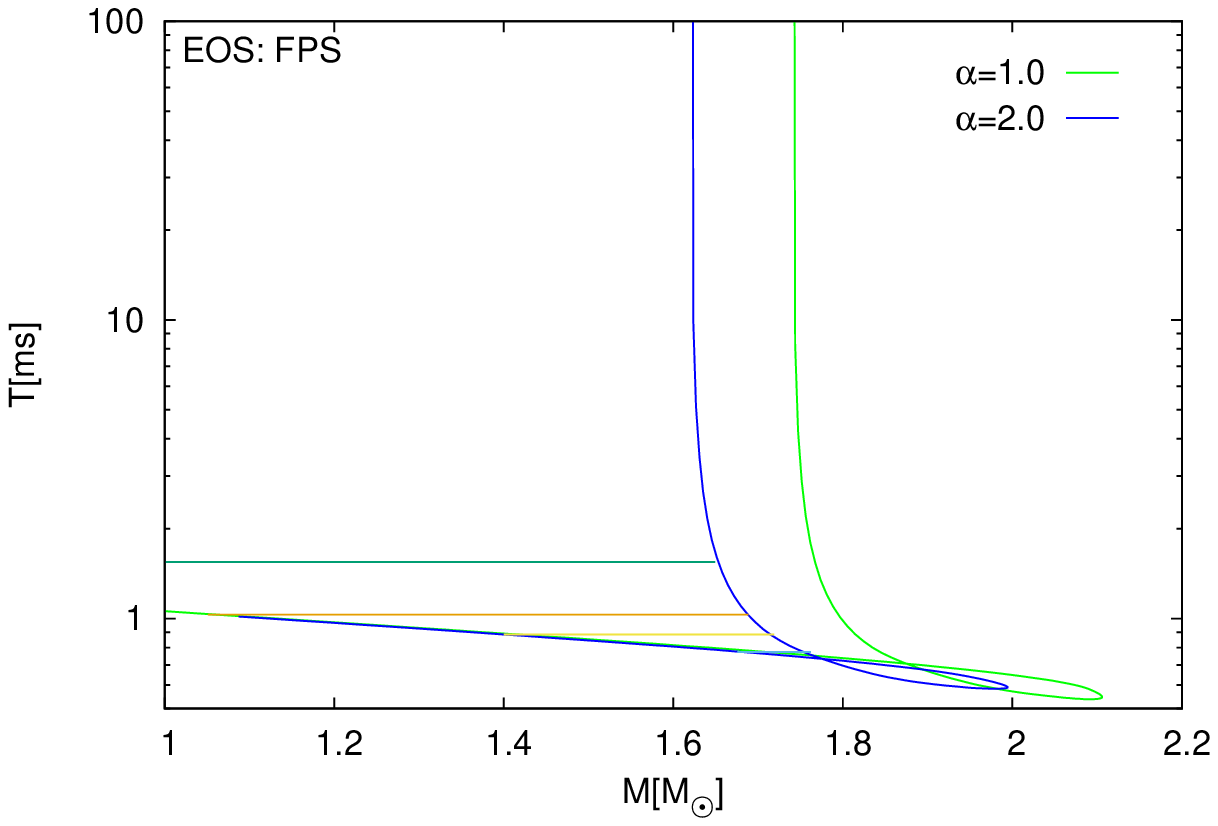}
(b)\includegraphics[height=.25\textheight, angle =0]{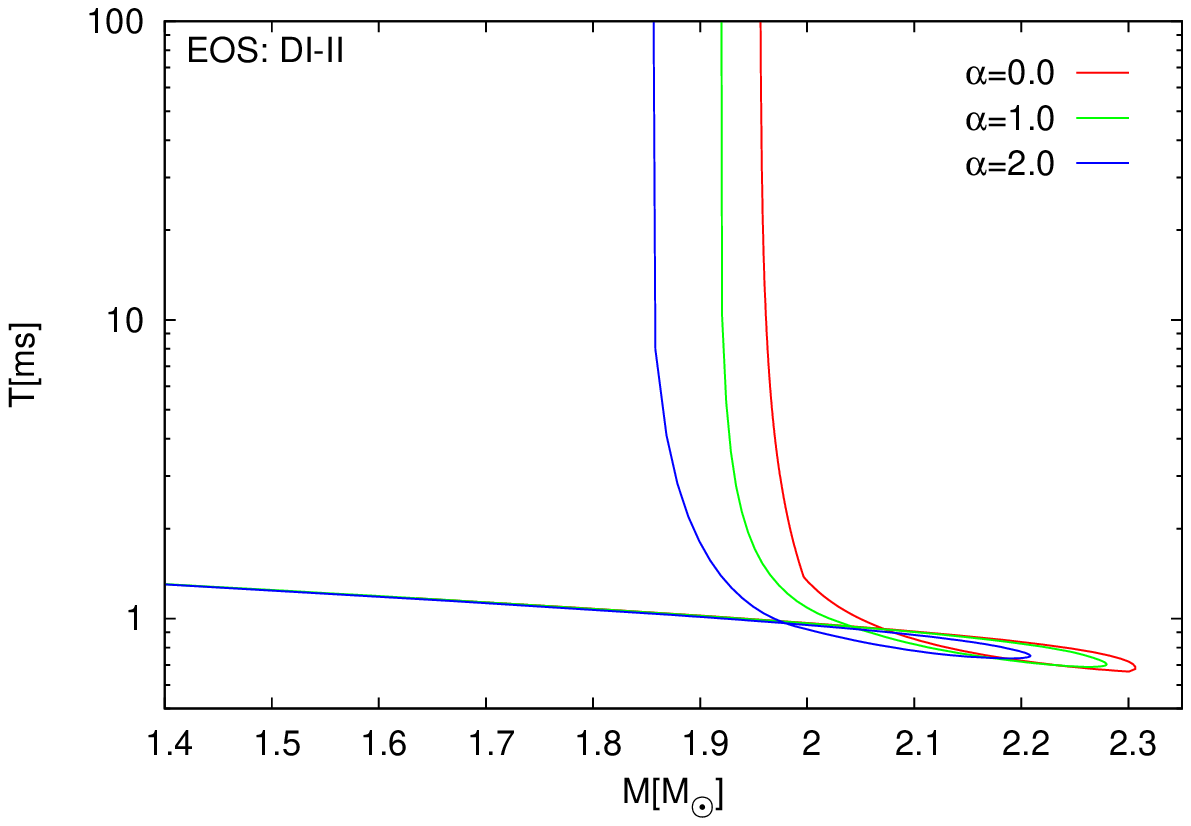}
}
\end{center}
\vspace{-0.5cm}
\caption{
(a) The rotation period $T$ is shown versus the mass $M$ for neutron stars 
in the physically relevant domain
for several values of the dimensionless angular velocity $\Omega$ 
for the EOS FPS.
The period $T$ is given in milliseconds,
the mass $M$ in units of the solar mass $M_\odot$.
 ($T=1.4$ ms corresponds to $f =716$ Hz.)
The GB coupling constant $\alpha$ has the values $\alpha=0$, 1 and 2.
The solid black lines represent 
the secular instability lines and the Keplerian sequences. 
(b) Same as (a) for the  EOS DI-II.
\label{fig9}
}
\end{figure}

The rotation periods $T$ of the known pulsars lie within the interval
$1.4$ ms $\le T \le 8.5$ s, i.e., the rotation period of the fastest known pulsar corresponds to only 1.4 ms.
We show the rotation period 
\begin{equation}
T=\frac{1}{f}=\frac{2 \pi}{\Omega}
\label{T}
\end{equation}
in units of milliseconds
versus the mass of neutron stars
in Fig.~\ref{fig9}.
Interestingly, close to the maximum mass, the period of neutron stars
along the Keplerian sequence is larger than the period along the
secular instability line. 
Therefore the boundary line exhibits a loop close to the maximum mass.

In contrast to neutron stars in GR, neutron stars in dEGB theory possess a 
scalar charge due to the presence of the dilaton field.
This dilaton charge $q$ has been defined in Eq.~(\ref{exdil}).
The presence of a scalar charge was addressed before in
\cite{Yagi:2011xp}, where it was shown, that
neutron stars do not possess a scalar charge in dEGB theory,
when the dilaton is coupled only linearly to the GB term.
In contrast, for the exponential coupling employed here
a small dilaton charge arises for neutron stars.

\begin{figure}[h!]
\begin{center}
\mbox{
(a)\includegraphics[height=.25\textheight, angle =0]{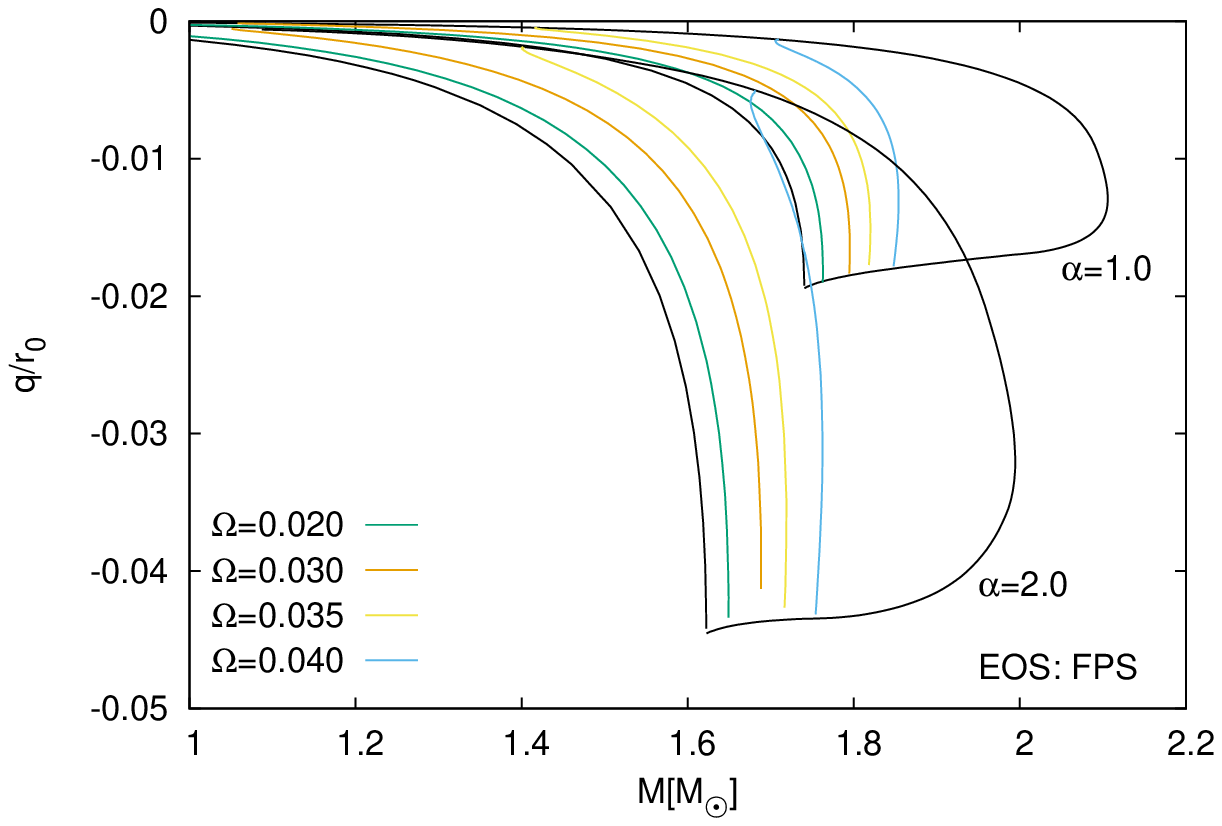}
(b)\includegraphics[height=.25\textheight, angle =0]{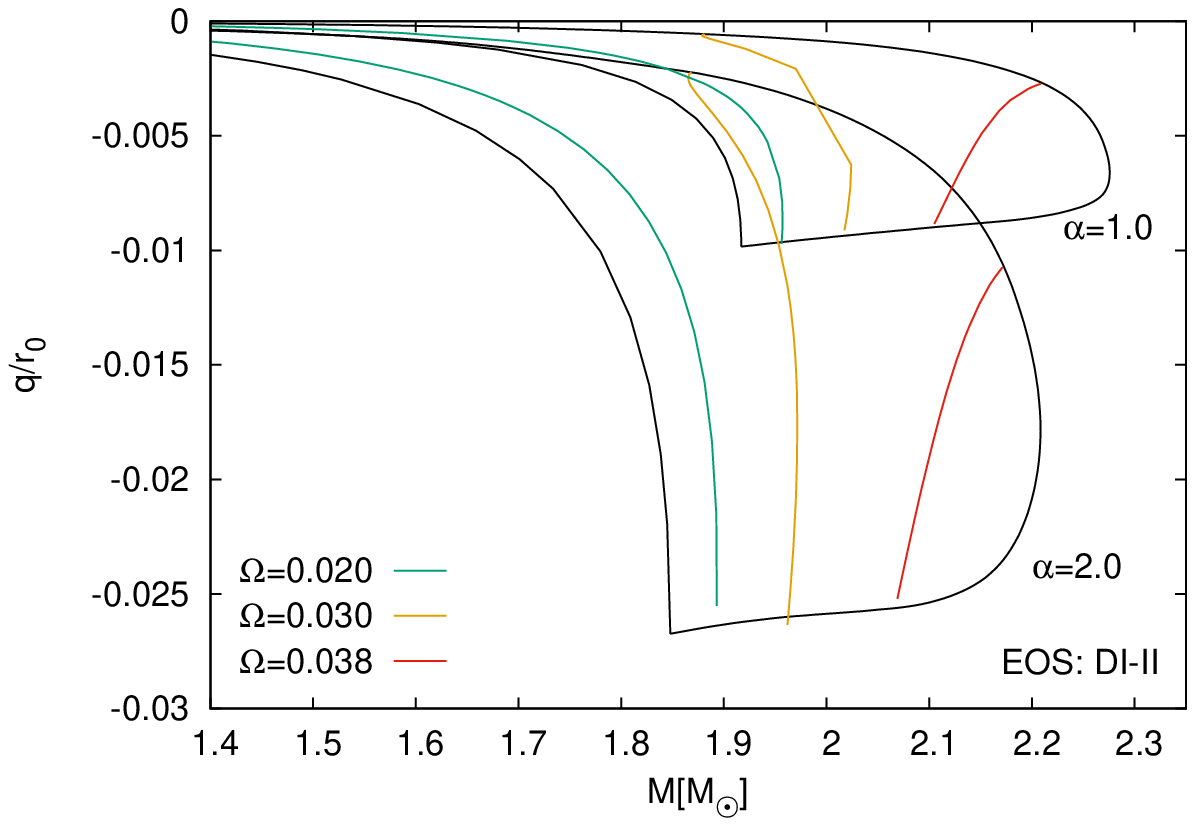}
}
\end{center}
\vspace{-0.5cm}
\caption{
(a) The dilaton charge $q$ is shown versus the mass $M$ for neutron stars 
in the physically relevant domain
for several values of the dimensionless angular velocity $\Omega$ 
for the EOS FPS. ($\Omega=0.01$ corresponds to $f =323$ Hz.)
The dilaton charge is scaled with
$r_0$, the mass $M$ is given in units of the solar mass $M_\odot$.
The GB coupling constant $\alpha$ has the values $\alpha=1$, and 2.
The solid black lines represent the static sequences,
the secular instability lines and the Keplerian sequences. 
(b) Same as (a) for the  EOS DI-II.
\label{fig7}
}
\end{figure}

In Fig.~\ref{fig7} we show the dilaton charge $q$ 
as a function of the mass $M$ for
GB coupling constants $\alpha=1$ and 2.
For neutron stars of small masses the magnitude of the scalar charge remains small.
It assumes its maximal value for static neutron stars in the stability limit.
As expected the magnitude of the scalar charge is strongly related to the
coupling parameter $\alpha$. 
For neutron stars obtained with the EOS FPS the magnitude of the 
scalar charge reaches larger values 
than for neutron stars obtained with the EOS DI-II.

\subsection{Moment of Inertia and Quadrupole Moment}

\begin{figure}[h!]
\begin{center}
\mbox{
(a)\includegraphics[height=.25\textheight, angle =0]{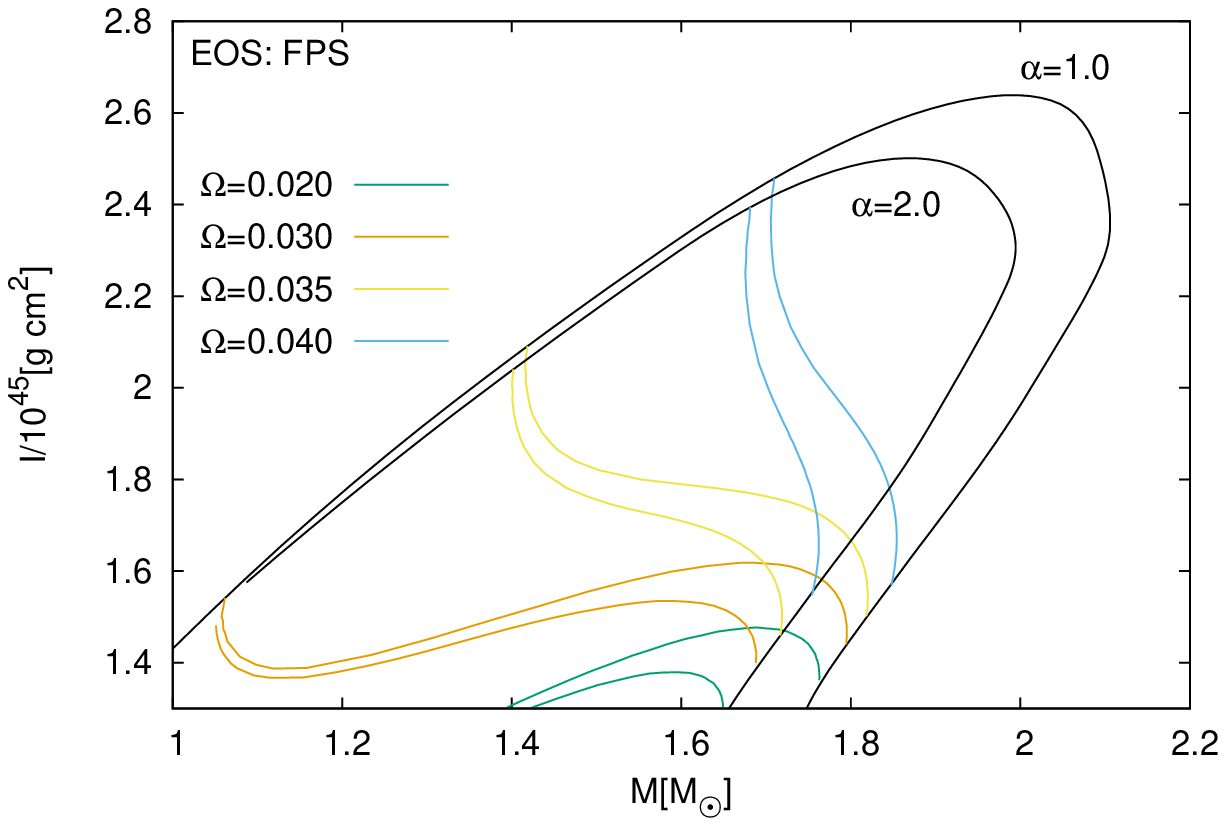}
(b)\includegraphics[height=.25\textheight, angle =0]{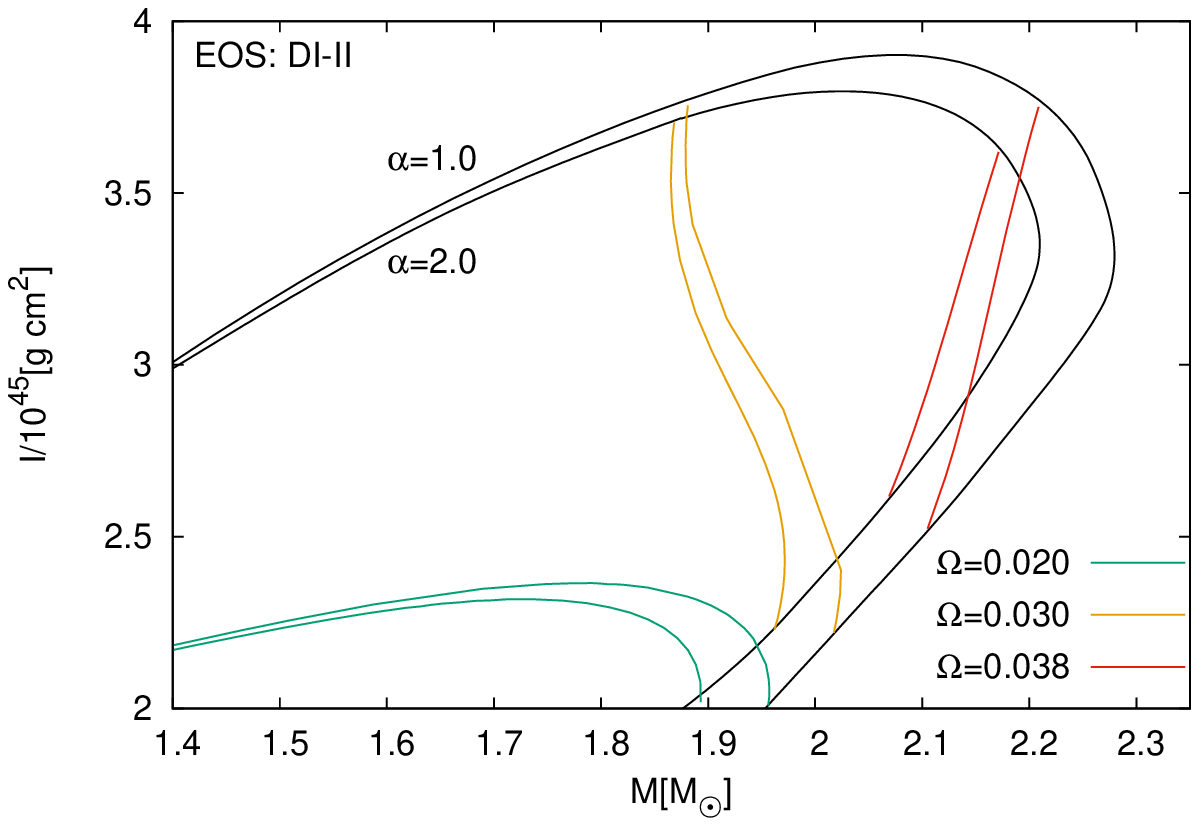}
}
\end{center}
\vspace{-0.5cm}
\caption{
(a) The moment of inertia $I$ is shown versus the mass $M$ for neutron stars 
in the physically relevant domain
for several values of the dimensionless angular velocity $\Omega$ 
for the EOS FPS. ($\Omega=0.01$ corresponds to $f =323$ Hz.)
The moment of inertia is given in units of $10^{45}$ $\rm g\, cm^2$,
the mass $M$ in units of the solar mass $M_\odot$.
The GB coupling constant $\alpha$ has the values $\alpha=1$, and 2.
The solid black lines represent
the secular instability lines and the Keplerian sequences. 
(b) Same as (a) for the  EOS DI-II.
\label{fig6I}
}
\end{figure}

We now turn to the moment of inertia $I$ of the rotating neutron stars
which is a very important physical quantity in the analysis of pulsars.
It can be obtained from the ratio of the angular momentum $J$ and the
angular velocity $\Omega$ 
\begin{equation}
I = \frac{J}{\Omega} .
\label{momin}
\end{equation}
In Fig.~\ref{fig6I} we exhibit the moment of inertia $I$ of neutron stars.
Note, that we do not give the moment of inertia obtained for slow rotation,
since we did not redo the perturbative calculations of \cite{Pani:2011xm}.
The figure therefore contains only the rapidly rotating sequences including 
the Keplerian sequence and the secular instability line.

The quadrupole moment $Q$ can be extracted from the asymptotic expansions of the
metric and the dilaton field, as given in Eq.~(\ref{Q}) (see Appendix A for a
brief derivation).
The static neutron stars are spherically symmetric, so their quadrupole
moment vanishes. For the sequences at fixed angular velocity $\Omega$
the quadrupole moment increases monotonically from the secular 
instability line to the Keplerian limit, where the star is maximally deformed.

\begin{figure}[h!]
\begin{center}
\mbox{
(a)\includegraphics[height=.25\textheight, angle =0]{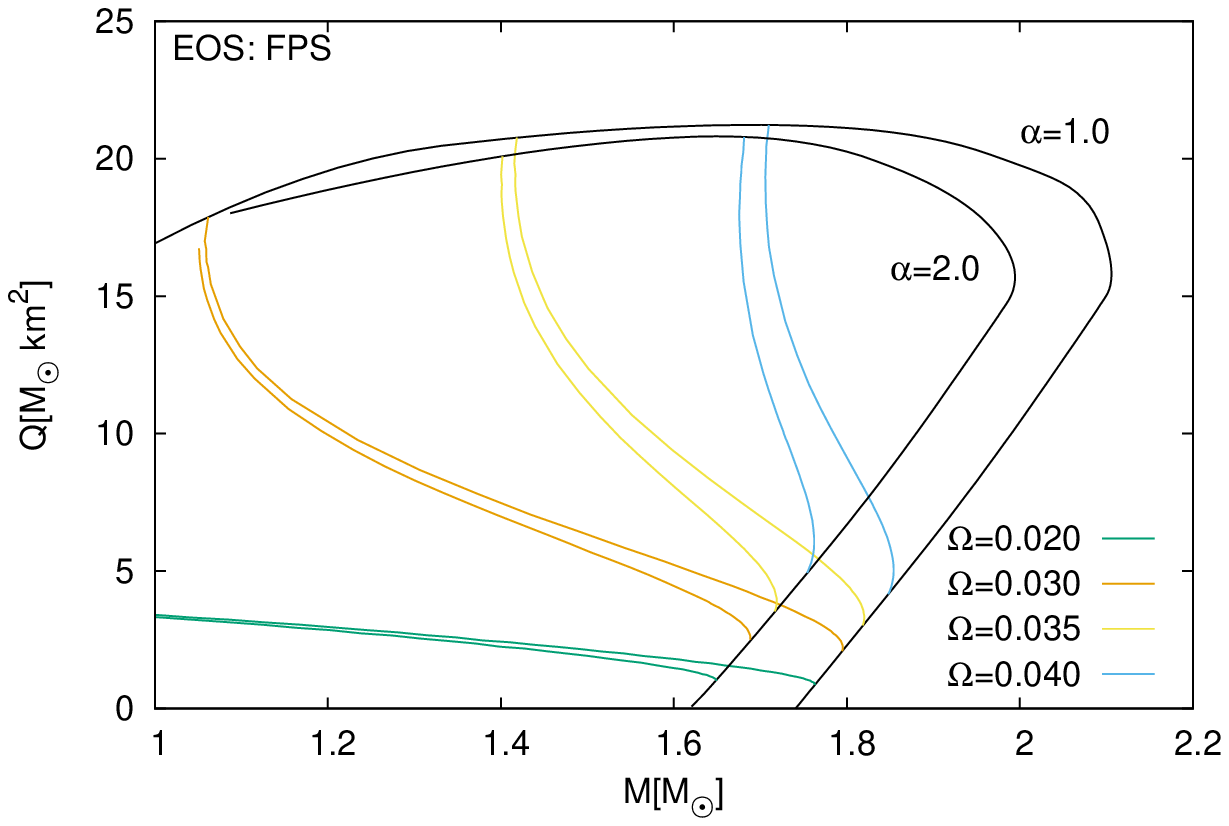}
(b)\includegraphics[height=.25\textheight, angle =0]{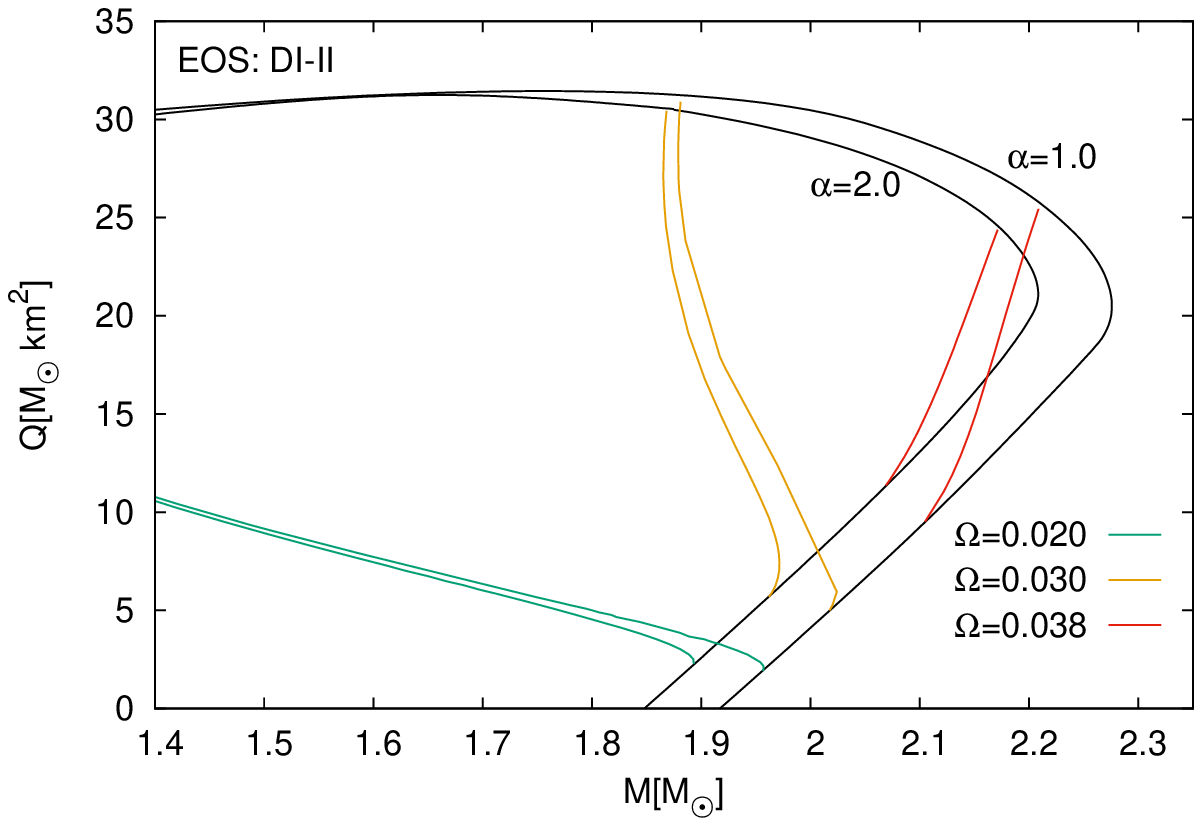}
}
\end{center}
\vspace{-0.5cm}
\caption{
(a) The quadrupole moment $Q$ is shown versus the mass $M$ for neutron stars 
in the physically relevant domain
for several values of the dimensionless angular velocity $\Omega$ 
for the EOS FPS. ($\Omega=0.01$ corresponds to $f =323$ Hz.)
The quadrupole moment is given in units of the solar mass times square
kilometers, $M_\odot$ km$^2$,
the mass $M$ in units of the solar mass $M_\odot$.
The GB coupling constant $\alpha$ has the values $\alpha=1$, and 2.
The solid black lines represent
the secular instability lines and the Keplerian sequences. 
(b) Same as (a) for the  EOS DI-II.
\label{fig5}
}
\end{figure}

The quadrupole moment $Q$ is shown as a function of the mass $M$ in Fig.~\ref{fig5} for
$\alpha=1$ and 2. Note, that the version of the rns code, which we have used,
does not extract the necessary expression for the quadrupole moment.
We have therefore omitted the comparison of the quadrupole moment
with the GR values in the figure,
where the quadrupole moment is given in units of  $M_\odot$ km$^2$
and the mass $M$ in units of $M_\odot$.
We observe that there is little dependence on $\alpha$ along most of the
Keplerian sequence, while  along the secular instability line the quadrupole moment
is larger for larger values of $\alpha$, when compared at the same mass.

\begin{figure}[h!]
\begin{center}
\mbox{
(a)\includegraphics[height=.25\textheight, angle =0]{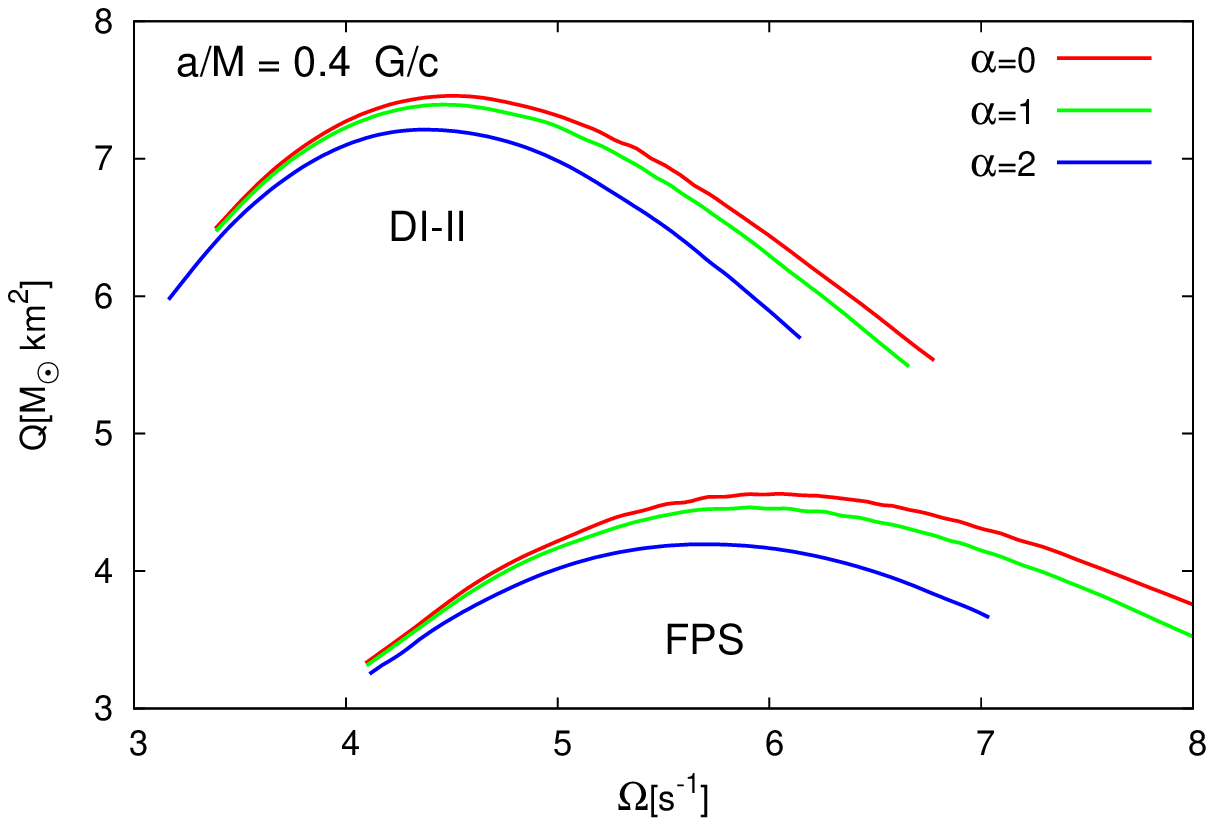}
(b)\includegraphics[height=.25\textheight, angle =0]{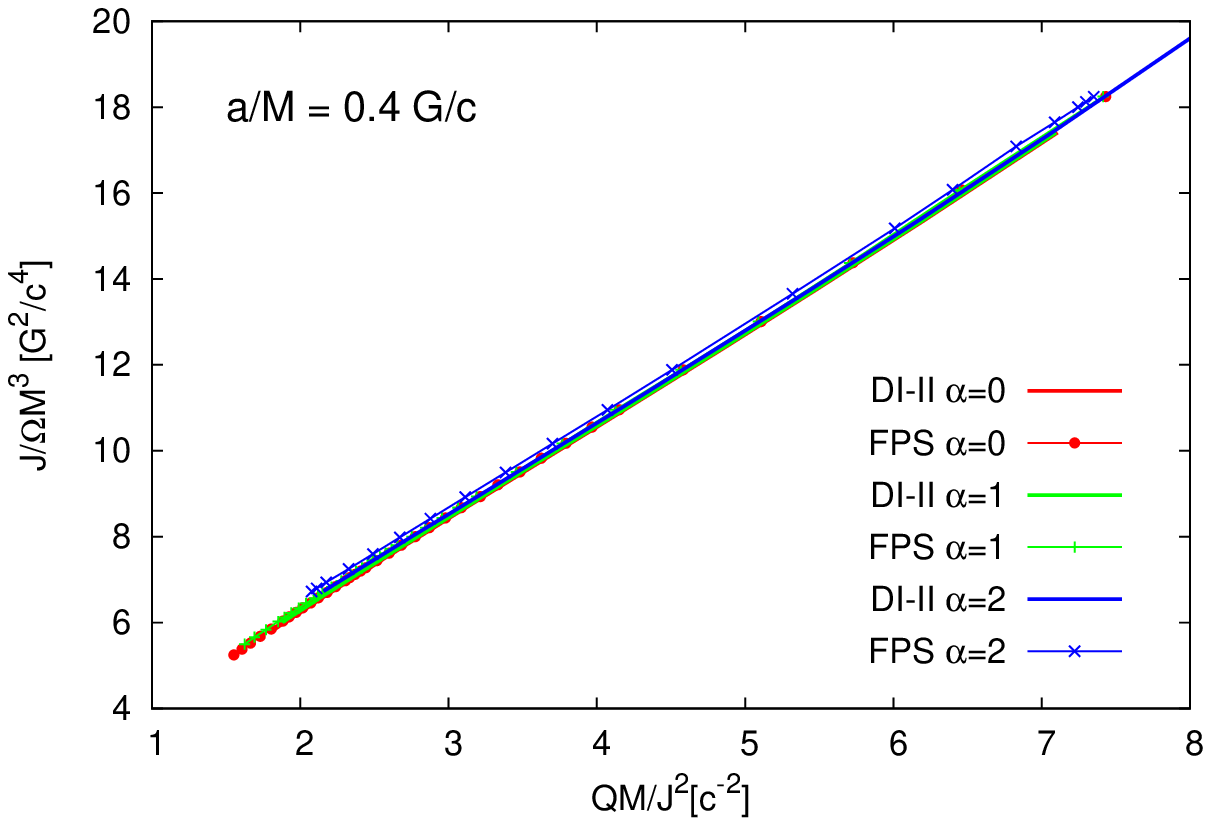}
}
\end{center}
\vspace{-0.5cm}
\caption{
(a) The quadrupole moment $Q$ is shown versus the angular velocity
$\Omega$ for neutron stars with fixed reduced dimensionless
angular momentum $a/M=0.4$.
The quadrupole moment is given in units of $M_\odot$ km$^2$,
the angular velocity $\Omega$ in units of Hz.
The GB coupling constant $\alpha$ has the values $\alpha=1$, and 2.
Employed are EOS DI-II and EOS FPS.
(b) The scaled moment of inertia $\hat{I}$ Eq.~(\ref{ihat}) versus
the scaled quadrupole moment $\hat{Q}$ Eq.~(\ref{qhat})
for the same set of solutions.
\label{fig6}
}
\end{figure}

Let us now consider the quadrupole moment for a different set of sequences
of neutron stars, where we fix the reduced dimensionless
angular momentum $a/M$, Eq.~(\ref{aj}), and vary the angular velocity $\Omega$.
We exhibit the quadrupole moment in units 
of $M_\odot \cdot {\rm km}^2$ as a function of 
the angular velocity in units of Hz in Fig.~\ref{fig6}(a) 
for the fixed value of $a/M=0.4$
and the values of the GB coupling constant $\alpha=0$, 1 and 2
for both EOSs employed, FPS and DI-II and FPS \cite{Kleihaus:2014lba}.
Obviously, the different EOSs give rise to rather different values 
for the quadrupole moment $Q$ for the same values 
of the angular velocity $\Omega$.
We also observe a pronounced dependence on $\alpha$ for the larger
angular velocities.

Clearly, the physical properties of neutron stars 
possess typically a pronounced dependence on the chosen EOS.
However, in recent years much effort has concentrated
on the study of quantities, which are 
independent of the EOS, or better, almost independent of the EOS.
A prominent example are the so-called $I$-Love $Q$ relations for neutron stars
which represent universal relations holding between the scaled moment
of inertia, the Love number, and the scaled quadrupole moment
in Einstein gravity \cite{Yagi:2013bca,Yagi:2014bxa,Yagi:2014qua}.

These relations have been considered first for the slowly rotating case 
\cite{Yagi:2013bca,Yagi:2014bxa,Yagi:2014qua}.
For instance, the $I$-$Q$ relation between 
the scaled moment of inertia $\hat I$
\begin{equation}
\label{ihat}
\hat{I} = \frac{J}{\Omega M^3} \frac{c^4}{G^2}
\end{equation}
and the scaled quadrupole moment $\hat{Q}$
\begin{equation}
\label{qhat}
\hat{Q} = \frac{QM}{J^2} c^2
\end{equation}
has yielded a  mean square fit for the function $\hat I(\hat Q)$, 
from which the values of any of the large number of different EOSs employed
differ by less than one percent \cite{Yagi:2013bca}.

For the generalization to rapidly rotating neutron stars,
it has turned out, that one needs to consider the relation $\hat I(\hat Q)$
at fixed values of the reduced dimensionless angular momentum $a/M$
in order to obtain near EOS independence
\cite{Chakrabarti:2013tca,Doneva:2013rha}.
This is the reason, that we have considered the fixed $a/M$ sequences in
Fig.~\ref{fig6}.

Let us now address the $I$-$Q$ relation for dEGB theory.
To this end we exhibit the scaled moment of inertia
$\hat{I}$ versus the scaled quadrupole moment $\hat{Q}$
in Fig.~\ref{fig6}(b) 
for a fixed value of $a/M=0.4$ both for the 
EOS DI-II and the EOS FPS \cite{Kleihaus:2014lba}. 
Clearly, the dependence of the $\hat{I}$-$\hat{Q}$ relation on 
the equation of state (for fixed $\alpha$) is very weak,
although it increases slightly with increasing $\alpha$.
Thus the dEGB theory possesses basically the same 
universal $I$-$Q$ relation as GR.
Similar findings were obtained in STT \cite{Doneva:2014faa}
(see also \cite{Berti:2015itd}).


\subsection{Deformation and Shape}

\begin{figure}[h!]
\begin{center}
\mbox{
(a)\includegraphics[height=.25\textheight, angle =0]{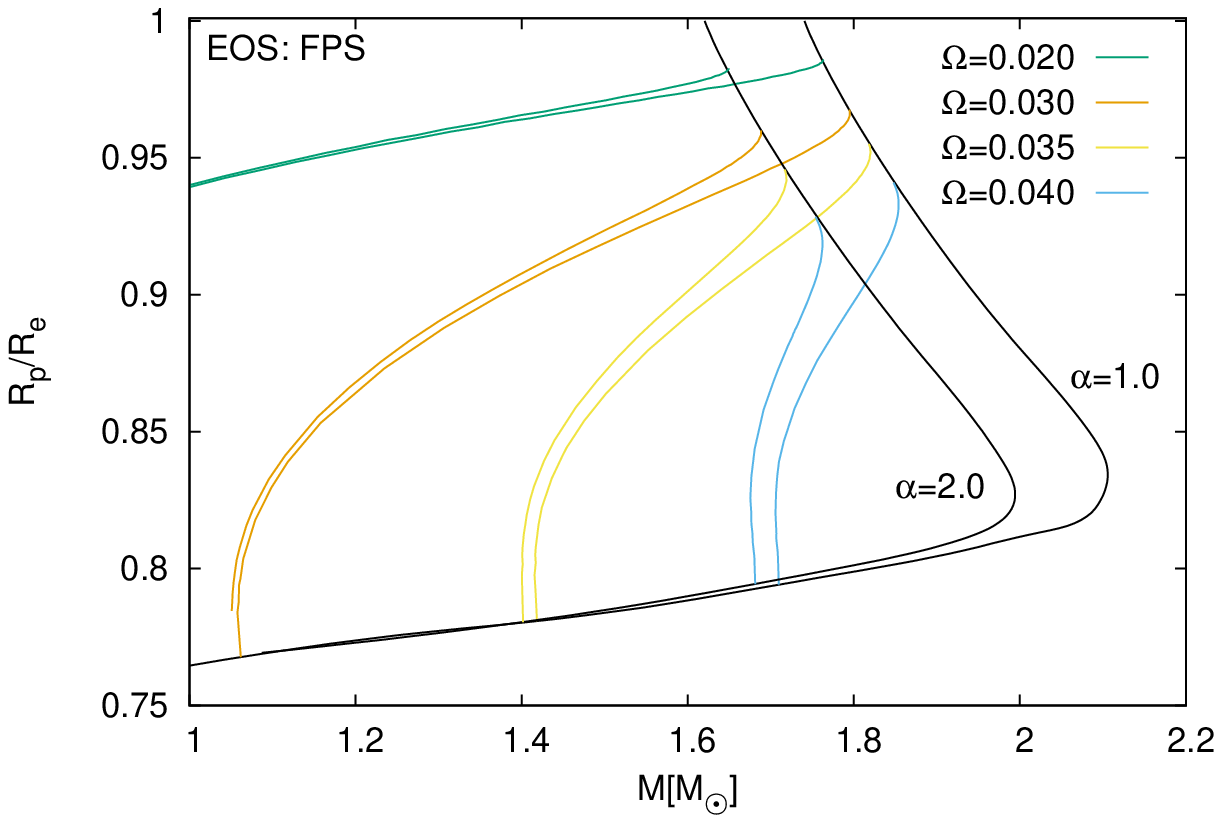}
(b)\includegraphics[height=.25\textheight, angle =0]{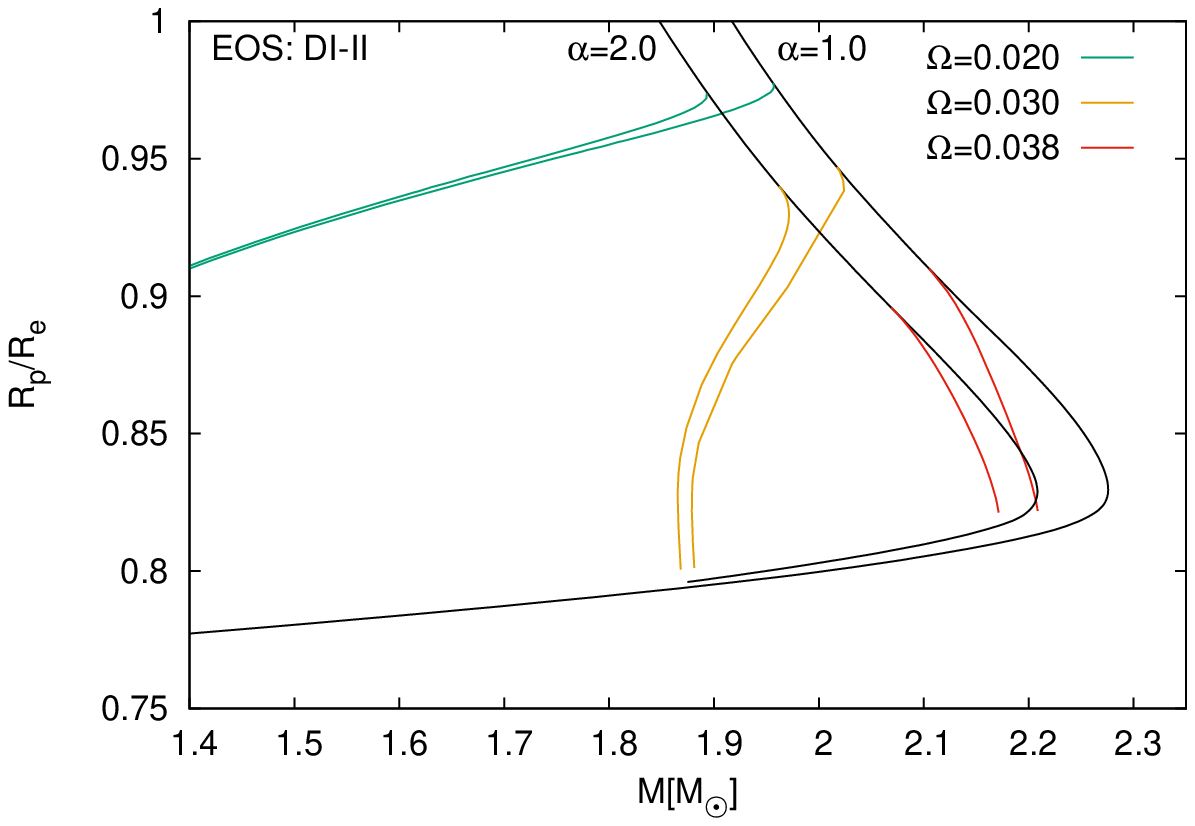}
}
\end{center}
\vspace{-0.5cm}
\caption{
(a) The ratio of the polar to the equatorial radius, $R_p/R_e$, 
 is shown versus the mass $M$ for neutron stars 
in the physically relevant domain
for several values of the dimensionless angular velocity $\Omega$ 
for the EOS FPS. ($\Omega=0.01$ corresponds to $f =323$ Hz.)
The mass $M$ is given in units of the solar mass $M_\odot$.
The GB coupling constant $\alpha$ has the values $\alpha=1$, and 2.
The solid black lines represent
the secular instability lines and the Keplerian sequences. 
(b) Same as (a) for the  EOS DI-II.
\label{fig10}
}
\end{figure}

Let us finally address the deformation and the shape 
of rapidly rotating neutron stars.
Since the centrifugal forces deform the neutron stars, 
their equatorial radius $R_e$
increases and the neutron stars flatten. To give an invariant account of this
deformation we consider the ratio of the polar radius $R_p$ to the 
equatorial radius $R_e$, defined in Eqs.~(\ref{r_p}) and (\ref{r_e}),
respectively.
In Fig.~\ref{fig10} we show this ratio 
versus the mass for $\alpha=1$ and 2 for both EOSs employed.
As expected the deformation is strongest along the Keplerian sequence,
when the star gets unstable with respect to losing mass.

\begin{figure}[h!]
\begin{center}
\mbox{
(a)\includegraphics[height=.25\textheight, angle =0]{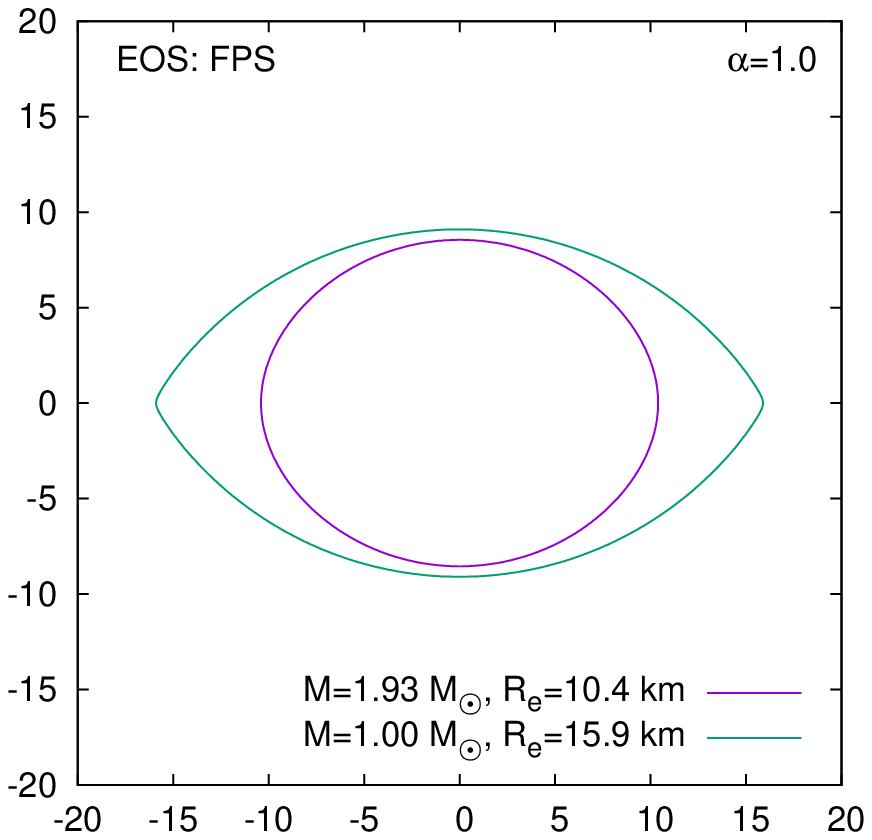}
(d)\includegraphics[height=.25\textheight, angle =0]{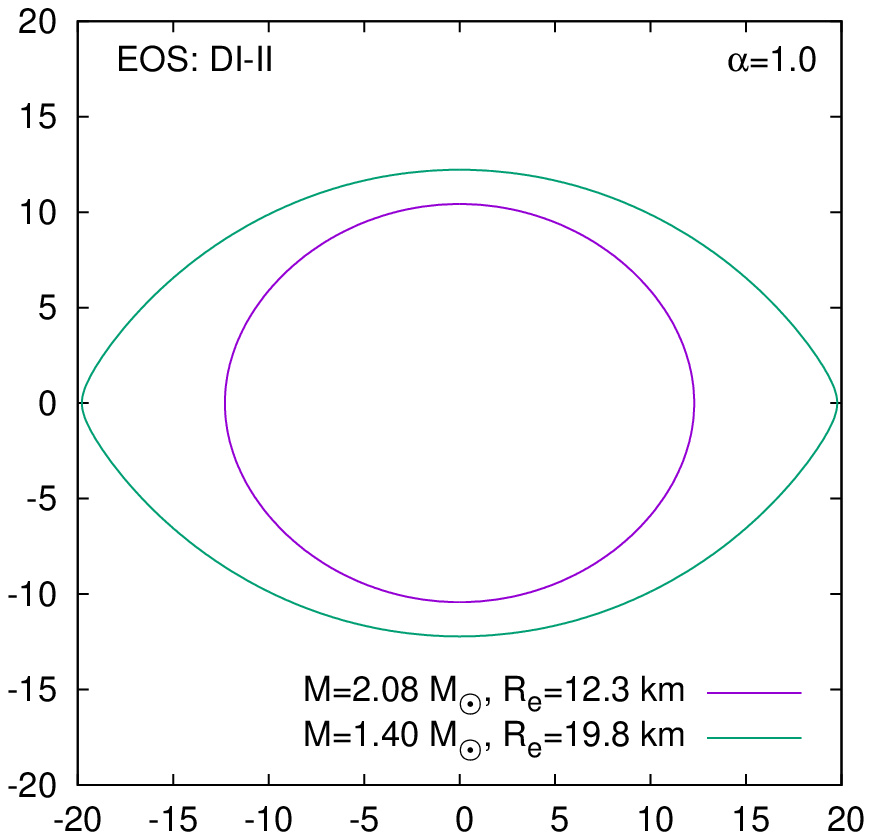}
}
\end{center}
\vspace{-0.5cm}
\caption{
(a) Isometric embedding of the surface of a star 
close to the secular instability line
with mass $M=1.93 M_\odot$ and equatorial radius $R_e=10.4$ km
and of a star close to the Kepler limit
with mass $M=1.00 M_\odot$ and equatorial radius $R_e=15.9$ km
for the EOS FPS and GB coupling constant $\alpha=1$.
(b) Analogous to (a) for stars with
mass $M=2.08 M_\odot$ and equatorial radius $R_e=12.3$ km and
mass $M=1.40 M_\odot$ and equatorial radius $R_e=19.8$ km
for the EOS DI-II.  
}
\label{fig11em}
\end{figure}

\begin{figure}[h!]
\begin{center}
\mbox{
(a)\includegraphics[height=.25\textheight, angle =0]{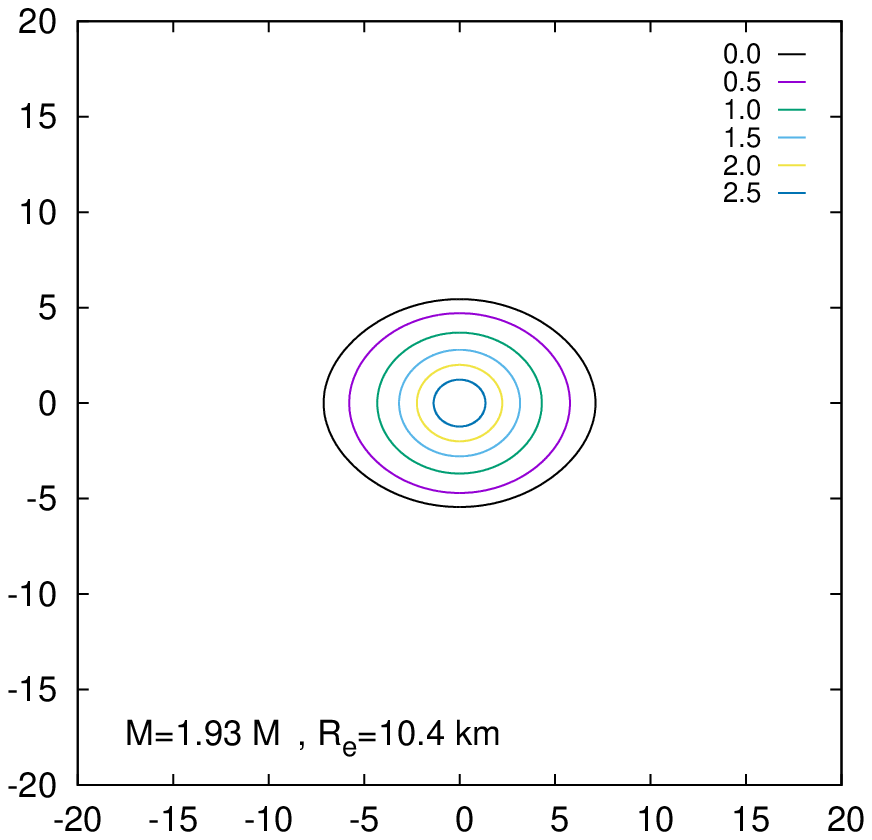}
(b)\includegraphics[height=.25\textheight, angle =0]{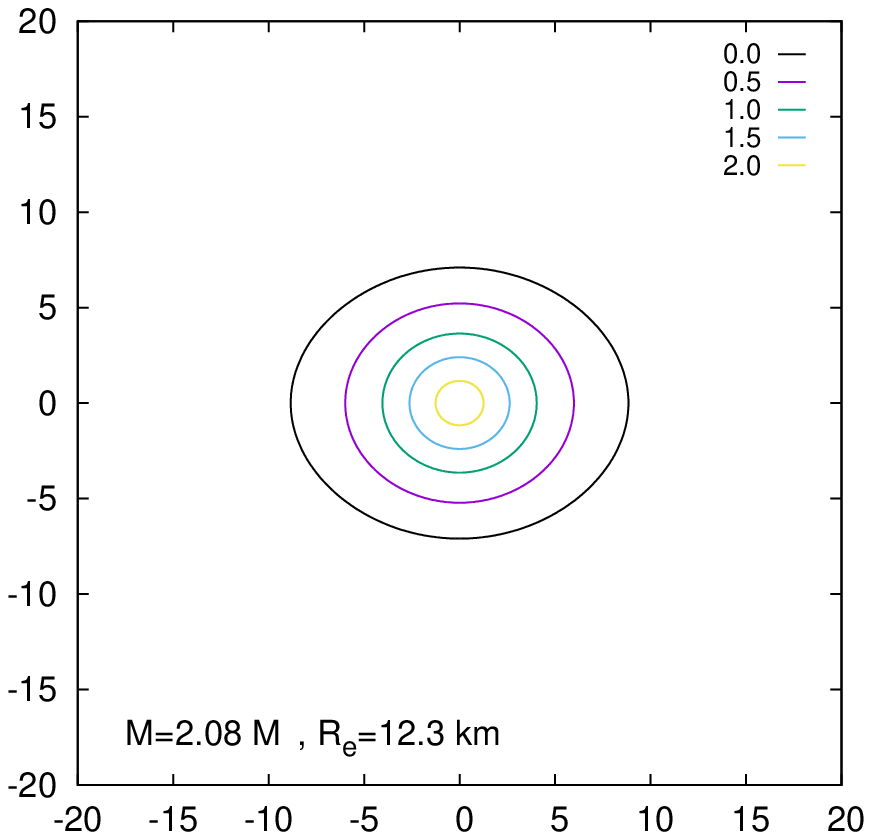}
}
\mbox{
(c)\includegraphics[height=.25\textheight, angle =0]{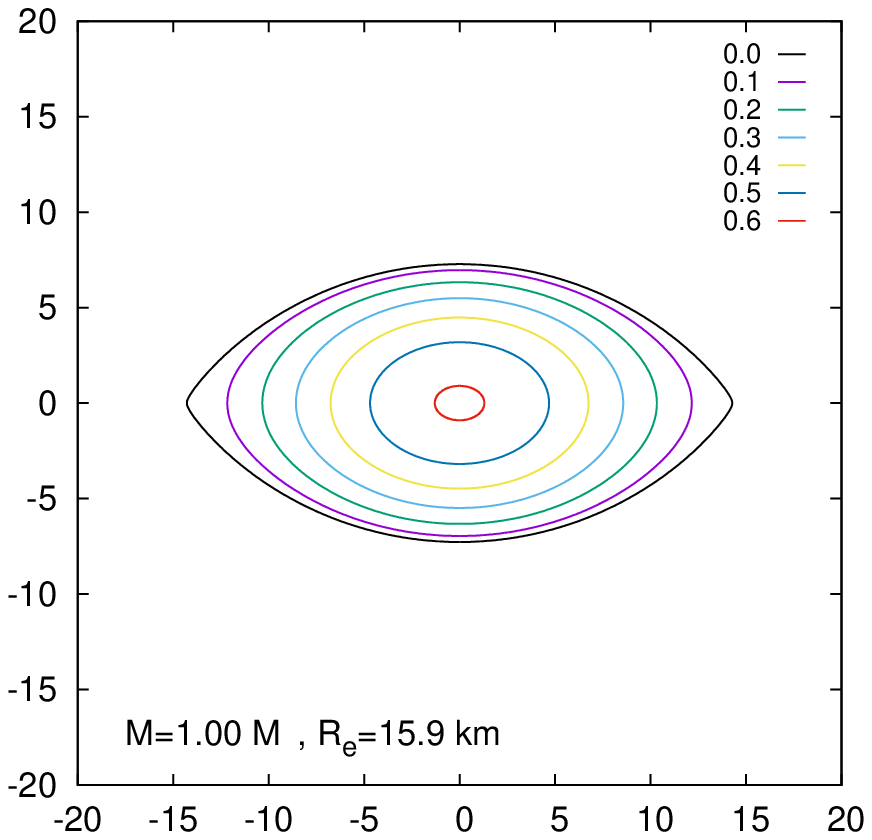}
(d)\includegraphics[height=.25\textheight, angle =0]{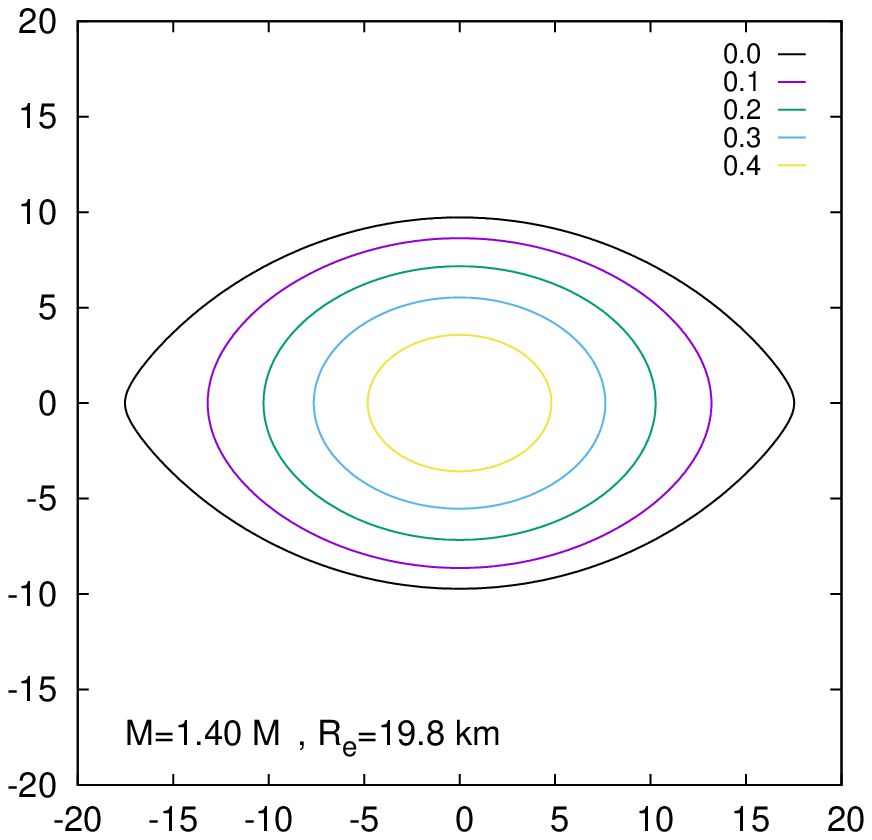}
}
\end{center}
\vspace{-0.5cm}
\caption{
(a) Isosurfaces of constant energy density $\epsilon/c^2$ 
in units of $10^{15}$ g/cm$^3$ for the large mass star of Fig.~\ref{fig11em}(a).
(b) Same as (a) for the large mass star of Fig.~\ref{fig11em}(b).
(c) Same as (a) for the lower mass star of Fig.~\ref{fig11em}(a).
(d) Same as (a) for the lower mass star of Fig.~\ref{fig11em}(b).
\label{fig11}
}
\end{figure}

To visualize the geometry of the surface of the neutron stars
we  calculate the isotropic embedding of the surface for several examples. 
We exhibit the isotropic embedding in 
Figs.~\ref{fig11em},
where we choose for each EOS one star close to the secular instability line
and one star close to the Kepler limit
and the GB coupling $\alpha=1$.
For the stars rotating close to the Kepler limit a cusp will develop
at the surface in the equatorial plane, when the limit is reached.

The distribution of the energy density $\epsilon$ is another physical quantity
of interest. We therefore exhibit contours of constant energy density
versus the coordinates $X= \pm r \sin\theta$ and $Z=  r \cos\theta$ 
($0 \leq \theta \leq \pi$) in Figs.~\ref{fig11}.
For better comparison, we choose the same neutron star solutions
for the contours of the energy density as for the isotropic embeddings.
The energy density $\epsilon/c^2$ contours are shown
in units of $10^{15}$ g/cm$^3$.
Note the higher central densities for the stars with the EOS FPS.
The GB coupling has the value $\alpha=1$.

\begin{figure}[h!]
\begin{center}
\hspace*{-2.5cm}
\mbox{
\includegraphics[height=.35\textheight, angle =0]{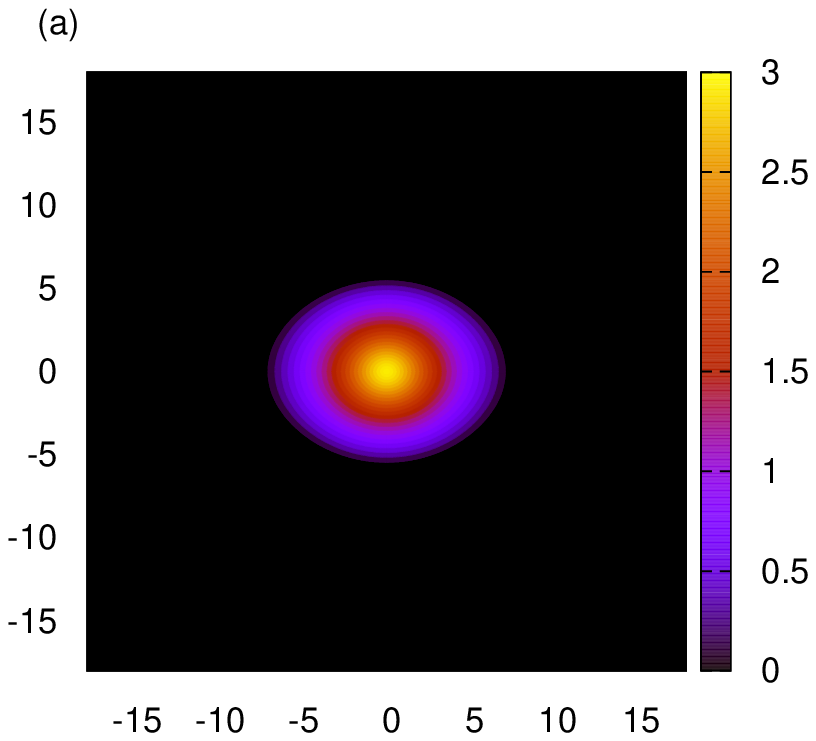}
\hspace*{-3cm}\includegraphics[height=.35\textheight, angle =0]{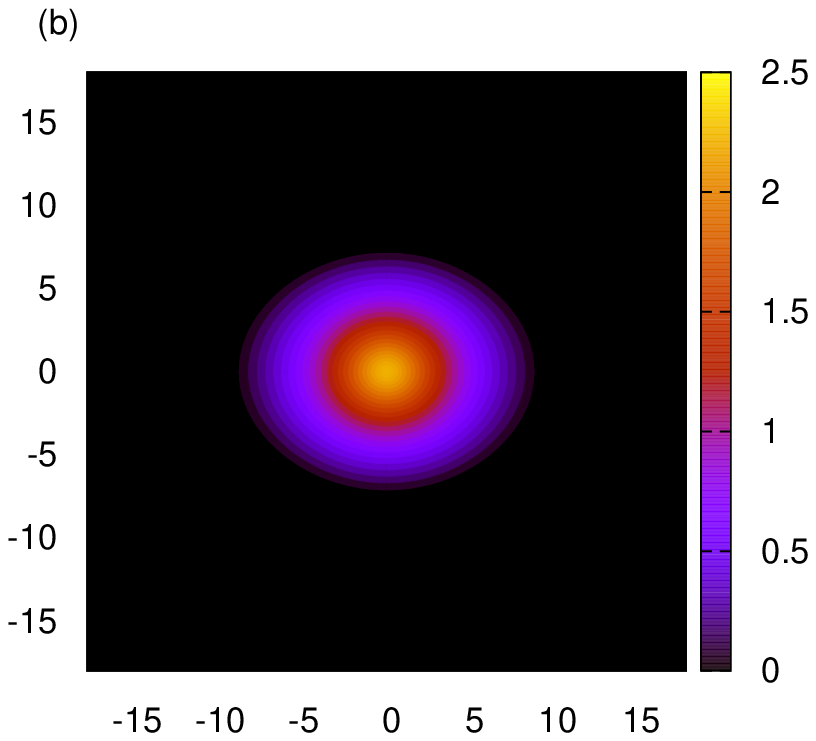}
}
\hspace*{-2.5cm}
\mbox{
\includegraphics[height=.35\textheight, angle =0]{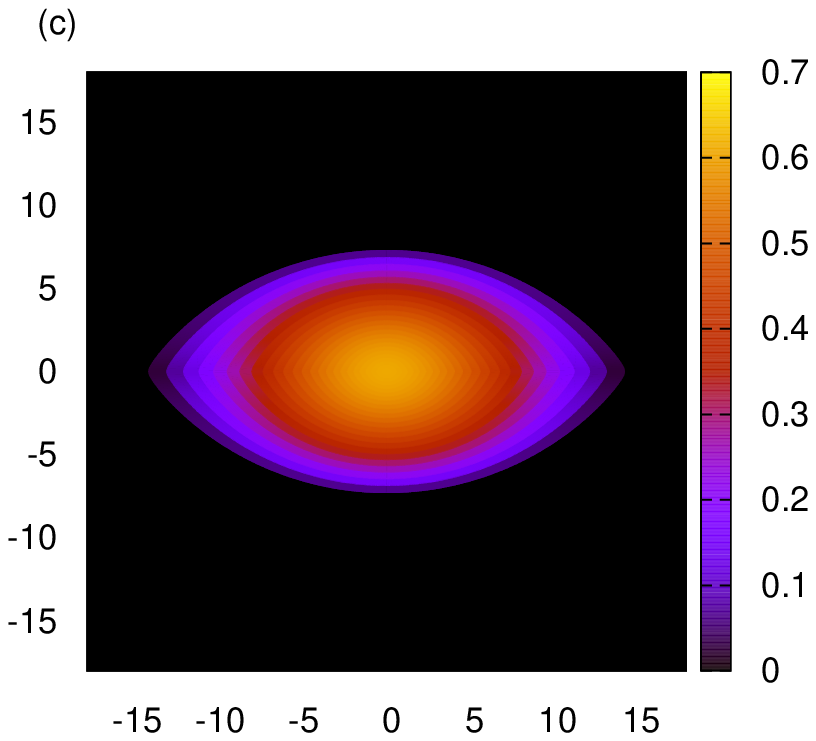}
\hspace*{-3cm}\includegraphics[height=.35\textheight, angle =0]{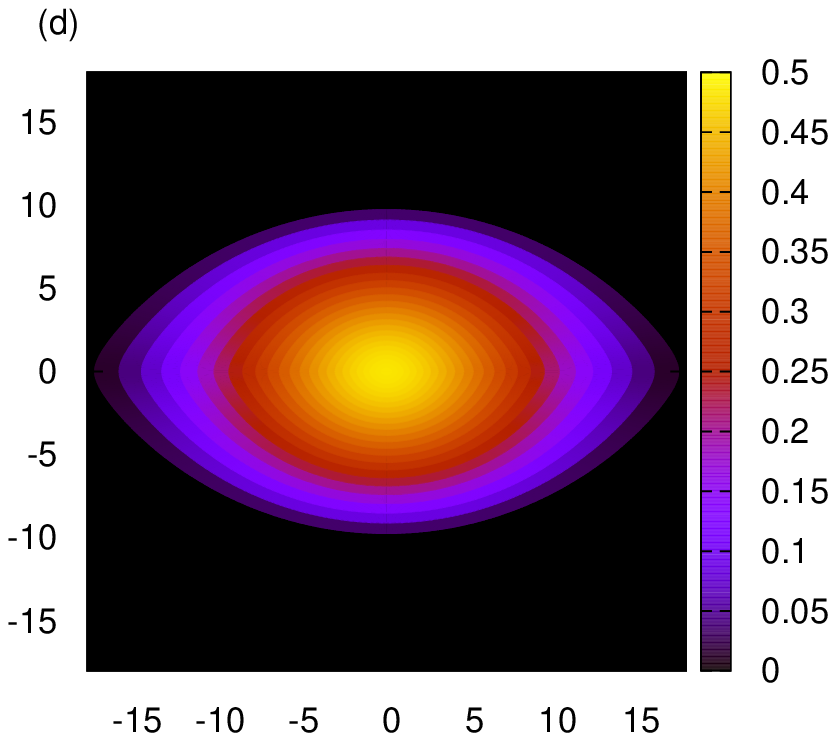}
}
\end{center}
\vspace{-0.5cm}
\caption{\label{fig12}
The energy density $\epsilon/c^2$ is shown 
in units of $10^{15}$ g/cm$^3$
for the examples of Figs.~\ref{fig11}.
}
\end{figure}

The inner structure of these neutron stars is illuminated further in 
Figs.~\ref{fig12}, 
where we exhibit colour encoded plots for the energy density for the same
set of examples and order of the figures as in Figs.~\ref{fig11}.

\section{Conclusions}

We have considered rapidly rotating neutron stars in dEGB theory,
a generalized model of gravity inspired by heterotic string theory.
Whereas static and slowly rotating neutron stars have been
studied before in this theory, we have here obtained for the
first time the full domain of physically relevant neutron stars
and its dependence on the GB coupling constant $\alpha$.

Delimited by the static and Keplerian sequences as well as by
the secular instability line, our results show that
this domain decreases as the
GB coupling constant $\alpha$ increases.
Employing dimensionless quantities we have chosen two values
for the dimensionless $\alpha$ (Eq.~(\ref{ab}),
$\alpha=1$ and 2, which are below the observational limit
($\alpha=12$) obtained
from binaries \cite{Yagi:2012gp}
and below the limit ($\alpha=3.4$) obtained from the study of
static neutron stars in dEGB theory \cite{Pani:2011xm}.
The latter limit arises from the fact that
if we were to increase $\alpha$ further, we would soon
encounter a genuine dEGB effect
known for black holes, wormholes and neutron stars
\cite{Kanti:1995vq,Pani:2011xm,Kanti:2011jz,Kleihaus:2011tg,Kleihaus:2015aje}:
somewhere along the sequences of neutrons stars
the solutions would stop existing, 
since the dilaton equation would no longer yield a real solution,
leading to a completely different type of boundary for the 
domain of rotating neutron stars
and also to a limit on $\alpha$ \cite{Pani:2011xm}.

For the EOS of the neutron stars we have selected two simple examples, the
EOS DI-II \cite{Diaz-Alonso:1985} and an approximation
to the EOS FPS \cite{Lorenz:1992zz,Haensel:2004nu},
both representing polytropic EOSs widely used before
in the calculation of neutron stars.
Since both EOSs do not yield a maximum mass of $2 M_\odot$
in the static case and for slow rotation, the present studies should
be extended to investigate further EOSs, yielding higher masses.
In the static case, this has already been achieved in
\cite{Blazquez-Salcedo:2015ets} for a set of eight realistic EOSs.
For the rapidly rotating case, however,
the current numerical scheme will need to be revised first
in order to achieve higher efficiency.
This also holds for a further increase of the GB coupling
constant $\alpha$.

Concerning the dependence of the physical properties of the neutron
stars on the GB coupling $\alpha$ we note that
the maximum masses are smaller for larger values of $\alpha$ while
the minimum radii are larger.
For the smaller masses and larger radii the Kepler limit
is (almost) independent of $\alpha$.
The compactness of the neutron stars decreases with increasing $\alpha$.

Besides the global charges
mass and angular momentum, neutron stars in dEGB theory
carry also a scalar charge, the dilaton charge.
This scalar charge arises only because of the exponential coupling
of the dilaton to the GB term. In contrast, for a linear coupling
the scalar charge would vanish \cite{Yagi:2011xp}.
For exponential coupling therefore scalar dipole radiation
would arise in a compact binary system 
whether or not one of the constituents is a black hole \cite{Yagi:2012gp}.

For the quadrupole moment of neutron stars
we have employed the definition of 
Geroch and Hansen \cite{Geroch:1970cd,Hansen:1974zz},
giving a brief derivation of the quadrupole moment in Appendix A.
The resulting expression corresponds to the
expression for Kerr-Newman black holes, when the
electric charge is replaced by the scalar charge.
The quadrupole moment exhibits a pronounced dependence
on the EOS and on the GB coupling.
The same is true for the moment of inertia.

The moment of inertia and the quadrupole moment
are known to exhibit a universal relation in GR
for slowly rotating neutron stars
\cite{Yagi:2013bca}
as well as for rapidly rotating neutron stars,
when the angular momentum is appropriately fixed
\cite{Chakrabarti:2013tca}.
We have shown, that in dEGB theory neutron stars
satisfy basically the same $I$-$Q$ relation as in GR,
presenting only a very weak dependence on the GB coupling $\alpha$.

Finally, we have addressed the deformation of the neutron stars
which arises because of the rotation.
For neutron stars close to the Kepler limit, one notices
the formation of a cusp at the surface in the equatorial plane.
Here mass shedding would occur when the Kepler limit is exceeded.
Clearly, this phenomenon is independent of the GB coupling
(as long as neutron stars rotating at the Kepler limit
exist in dEGB theory).

Turning now to future extensions of this work,
first of all a more efficient numerical code should be developed,
which should allow the study of rapidly rotating neutron stars
for a larger number of EOSs 
as well as the determination of the domain of existence
of neutron star solutions at large GB coupling.

From a string theoretical point of view, it would be very interesting
to include further terms into the action. 
Here, in particular, the effects of the Lorentz-Chern-Simons terms 
should be considered for rapidly rotating neutron stars
as well as the presence of Kalb-Ramond axions \cite{Campbell:1990ai}. 

\section*{Acknowledgements}

{We gratefully acknowledge discussions with
Tibault Damour, Norman G\"urlebeck and Eugen Radu
as well as support by the DFG
Research Training Group 1620 ``Models of Gravity''
and by the grant FP7, Marie Curie Actions, People,
International Research Staff Exchange Scheme (IRSES-606096).}


\appendix
\section{Quadrupole Moment}

Here we give a brief discussion of the derivation of the
quadupole moment for neutron stars in order to see
the effect of the dilaton and the GB term.
We employ the definitions of Geroch and Hansen
\cite{Geroch:1970cd,Hansen:1974zz},
and follow closely the later references
\cite{Hoenselaers:1992bm,Sotiriou:2004ud,Pappas:2012ns}.

Let us start by reviewing the definition of the quadrupole moment 
in \cite{Hoenselaers:1992bm}.
Let $\xi$ be a time-like Killing vector field on the space-time manifold
with metric $g$ and $\lambda$ the squared norm of $\xi$. We then define the 
metric $h$ on a 3-dimensional space by the projection
\begin{equation}
h = -\lambda g + \xi \otimes \xi\ .
\label{metrich}
\end{equation}
A 3-dimensional space $({\cal M},h)$ is called asymptotically flat 
if it can be conformally mapped to a manifold 
$(\tilde{\cal M},\tilde{h})$ with the properties
\begin{itemize}
\item[(i)] $\tilde{{\cal M}}= {\cal M}\cup \Lambda $, 
where $\Lambda \in \tilde{{\cal M}}$ 
\item[(ii)] $\left. \tilde{\Omega} \right|_\Lambda 
 = \nabla_i \left. \tilde{\Omega} \right|_\Lambda =0 $
            and 
$\nabla_i\nabla_j \left. \tilde{\Omega} \right|_\Lambda 
=\left. \tilde{h}_{ij}\right|_\Lambda$, 
where $\tilde{h}_{ij} =\tilde{\Omega}^2 h_{ij}$ .
\end{itemize}
In \cite{Hoenselaers:1992bm} the complex multipole tensors 
are defined recursively as follows,
\begin{eqnarray}
\tilde{\cal P}^{(0)} & = & \tilde{\Phi}  ,
\nonumber\\
\tilde{\cal P}_i^{(1)} & = & \partial_i\tilde{\Phi}  ,
\nonumber\\
\tilde{\cal P}_{i_1 \cdots i_{n+1}}^{(n+1)}
& = & {\cal C}\left[ \tilde{\nabla}_{i_{n+1}} \tilde{\cal P}_{i_1 \cdots i_{n}}^{(n)}
       -\frac{1}{2}n(2n-1)\tilde{R}_{i_1 i_2} \tilde{\cal P}_{i_3 \cdots i_{n+1}}^{(n-1)}
      \right] .
\label{cplxP}      
\end{eqnarray}
Here ${\cal C}$ denotes the symmetric trace-free part, 
$\tilde{R}_{ij}$ the Ricci tensor and 
$\tilde{\nabla}_i$ the covariant derivative on $(\tilde{\cal M},\tilde{h})$. 
$\tilde{\Phi}=\tilde{\Omega}^{-1/2}\Phi$,
where $\Phi$ is the complex mass potential.
Setting $n=1$ we find for the complex quadrupole 
\begin{equation}
\tilde{\cal P}_{ij}^{(2)}
 = {\cal C}\left[ \tilde{\nabla}_{j}\tilde{\nabla}_{i} \tilde{\Phi}
       -\frac{1}{2}\tilde{R}_{ij} \tilde{\Phi}
      \right] .
\label{quad1}     
\end{equation}

Note, that the line element on $\cal M$ in \cite{Hoenselaers:1992bm} is chosen in Weyl coordinates,
\begin{equation}
dh^2 = e^{2\gamma}\left(d\rho^2 +dz^2\right) + \rho^2 d\vphi^2  .
\label{dh2}
\end{equation}
The transformation
\begin{equation}
\bar{\rho} = \frac{\rho}{\rho^2+z^2}  , \ \ \ \ 
\bar{z}    = \frac{z}{\rho^2+z^2} \ 
\label{trans}
\end{equation}
leads to
\begin{equation}
dh^2 = \frac{1}{\bar{r}^4}\left[e^{2\gamma}\left(d\bar{\rho}^2 +d\bar{z}^2\right) 
+ \bar{\rho}^2 d\vphi^2\right] \ ,
\label{dh2t}
\end{equation}
with $\bar{r}^2 = \bar{\rho}^2 +\bar{z}^2$. 
Employing the conformal factor $\tilde{\Omega} = \bar{r}^2$ 
then leads to the line element on
 $\tilde{\cal M}$
\begin{equation} 
d\tilde{h}^2 = \tilde{\Omega}^2 dh^2 = 
e^{2\gamma}\left(d\bar{\rho}^2 +d\bar{z}^2\right) + \bar{\rho}^2 d\vphi^2 \ .
\label{dbarh2}
\end{equation}

Let us now turn to the 
EGBd solutions, which are obtained with the ansatz for the line element 
Eq.~(\ref{metric})
\begin{equation}
ds^2 = -e^{2\nu_0} dt^2 + e^{2(\nu_1-\nu_0)}
\left(e^{2\nu_2}\left[ dr^2+r^2 d\theta^2\right]
       +r^2 \sin^2\theta \left(d\varphi -\omega dt\right)^2\right)  
\label{ds2}       
\end{equation}
and consider its asymptotic behavior.
%
%
In the asymptotic region the metric and dilaton functions possess the expansion
(\ref{exnu0})-(\ref{exdil})
\begin{eqnarray}
\nu_0 & = & -\frac{M}{r} + \frac{D_1 M}{3r^3} - \frac{M_2}{r^3} P_2(\cos\theta) +{\cal O}(r^{-4}) ,
\label{Anu0}
\\
\nu_1 & = & \frac{D_1}{r^2} +{\cal O}(r^{-3}) ,
\label{Anu1}
\\
\nu_2 & = & -\frac{4 M^2+16 D_1+ q^2}{8 r^2}\sin^2\theta +{\cal O}(r^{-3}) ,
\label{Anu2}
\\
\omega & = & \frac{2 J}{r^3}  +{\cal O}(r^{-4}) ,
\label{Aom}
\\
\phi & = & \frac{q}{r}  +{\cal O}(r^{-2}) .
\label{Aphi}
\end{eqnarray}
We note that the function $\omega$ does not contribute to the qudrupole moment 
due to its fast decay. Consequently, we will neglect $\omega$
in the following. 
Comparison of the line elements (\ref{dh2}) and (\ref{ds2}) then yields
$\lambda = e^{2\nu_0}$ and 
\begin{eqnarray}
dh^2 & = & e^{2\gamma}\left(d\rho^2 +dz^2\right) + \rho^2 d\vphi^2
\label{dh22}\\
 & = &    e^{2(\nu_2+\nu_1)}\left( dr^2+r^2 d\theta^2\right)
          +e^{2\nu_1}r^2 \sin^2\theta d\vphi^2
\nonumber\\
 & = &    e^{2(\nu_2+\nu_1)}\left( d\hat{\rho}^2+d\hat{z}^2\right)
          +e^{2\nu_1}\hat{\rho}^2\theta d\vphi^2 ,
\label{dshath}
\end{eqnarray}
where we defined $\hat{\rho} = r \sin\theta$ and $\hat{z} = r \cos\theta$.

In the next step we transform to Weyl coordinates in the asymptotic region.
Comparision of (\ref{dh22}) and (\ref{dshath}) yields
\begin{eqnarray}
\rho & = & \hat{\rho}e^{\nu_1} = \hat{\rho}\left(1+\frac{D_1}{r^2}+{\cal O}(r^{-3})\right) ,
\label{eqrhow}\\
0    & = & \rho_{,\hat{\rho}} \rho_{,\hat{z}} + z_{,\hat{\rho}} z_{,\hat{z}} ,
\label{eqmix}\\
e^{2\gamma} & = & e^{2(\nu_2+\nu_1)} \left[\rho_{,\hat{\rho}}^2+z_{,\hat{\rho}}^2\right]^{-1} ,
\label{eqgam1}\\
e^{2\gamma} & = & e^{2(\nu_2+\nu_1)} \left[\rho_{,\hat{z}}^2+z_{,\hat{z}}^2\right]^{-1} .
\label{eqgam2}
\end{eqnarray}
We solve Eq.~(\ref{eqmix}) by the ansatz $z_{,\hat{\rho}} = \rho_{,\hat{z}}$
$z_{,\hat{z}} = -\rho_{,\hat{\rho}}$. Integration then yields 
\begin{equation}
z = \hat{z}\left(-1+\frac{D_1}{r^2}+{\cal O}(r^{-3})\right)  , 
\label{eqzw}
\end{equation}
where the integration constant has been set to zero.
Note, that with this ansatz Eqs.~(\ref{eqgam1}) and (\ref{eqgam2}) are identical.

Inversion of the above relations yields the coordinates $\hat{\rho}$, $\hat{z}$ as 
functions of the Weyl coordinates $\rho$, $z$,
\begin{equation}
\hat{\rho}=\rho\left( 1-\frac{D_1}{r'^2}+{\cal O}(r'^{-3})\right) \ , \ \ \ \ 
\hat{z}   = z  \left(-1-\frac{D_1}{r'^2}+{\cal O}(r'^{-3})\right) \ , 
\label{invco}
\end{equation}
with $r'^2= \rho^2 + z^2$.
Substitution in the expansion Eqs.~(\ref{Anu0})-(\ref{Anu2}) and 
$\left[\rho_{,\hat{\rho}}^2+z_{,\hat{\rho}}^2\right]$ then yields
\begin{equation}
\gamma = -\frac{\rho^2 \left(4 M^2 +q^2\right)}{8 r'^4} +{\cal O}(r'^{-4}) \ .
\label{gamWC}
\end{equation}
In these coordinates we find for the mass potential $\Phi$
\begin{equation}
\Phi = \frac{\lambda^2 -1}{4\lambda} = 
-\frac{1}{r'}\left(M+\frac{2M^3}{3r'^2}
                    -\left[\frac{M_2}{2}-\frac{2}{3}M D_1\right]\frac{\rho^2-2z^2}{r'^4}
		    +{\cal O}(r'^{-3})\right)  .
\label{potWC}
\end{equation}
Next we use the conformal mapping 
\begin{equation}
\rho = \frac{\bar{\rho}}{\bar{r}^2}  , \ \ \ \ 
z    = \frac{\bar{z}}{\bar{r}^2}   , \ \ \ \ 
\label{transinv}
\end{equation}
to find $\gamma$ and the mass potential 
$\tilde{\Phi}= \frac{1}{\bar{r}} \Phi = r' \Phi $
on the 3-dimensional manifold $\tilde{\cal M}$,
\begin{eqnarray}
\gamma & = & -\frac{\bar{\rho}^2 \left(4 M^2 +q^2\right)}{8} +{\cal O}(\bar{r}^4)  ,
\label{tgamWC}\\
\tilde{\Phi} & = &
-\left(M+\frac{2M^3}{3}\bar{r}^2
                    -\left[\frac{M_2}{2}-\frac{2}{3}M D_1\right]\left(\bar{\rho}^2-2\bar{z}^2\right)
		    +{\cal O}(\bar{r}^3)\right)  .
\label{tpotWC}
\end{eqnarray}

Since by now everything is expressed in the appropriate coordinates 
we can apply Eq.~(\ref{quad1}) 
to compute the $\bar{z}\bar{z}$ component of the 
quadrupole tensor at $\Lambda=0$, i.e., $\bar{r}=0$. 
The scalar quadrupole moment $Q$ is then given by Eq.~(\ref{Q})
\begin{equation}
Q= \frac{1}{2}\tilde{\cal P}_{zz}^{(2)}(0) 
=  -M_2 +\frac{4}{3}\left[\frac{1}{4}+\frac{D_1}{M^2}+\frac{q^2}{16M^2}\right] M^3 .
\label{Qapp}
\end{equation}
Note, that in the vacuum limit, when the dilaton charge vanishes,
the quadrupole  moment $Q$ is the same as in \cite{Pappas:2012ns} 
(up to an overall sign),
with $b=D_1/M^2$.
On the other hand, in the static limit the solution is spherically symmetric.
In this case $M_2=0$ and $4M^2 + 16 D_1 + q^2 =0$. Consequently, the quadrupole moment 
vanishes in the static limit.

We further note, that EGBd theory and Einstein-Maxwell
theory have the same expansion as far as the lower order terms are concerned.
First, there is no contribution from the Gauss-Bonnet term 
since it decays sufficiently fast.
Second, the dilaton field enters only via the Coulomb-like term $\frac{q}{r}$, 
which coincides with the analogous term in Einstein-Maxwell theory in the 
lowest order of the expansion. 
Therefore, since EGBd theory and Einstein-Maxwell theory 
then describe the same spacetime (up to higher order terms), 
the expression for the quadrupole moment also coincides for both theories.

\end{document}